                \documentclass{aa}
\usepackage{graphicx}
\usepackage{txfonts}

\usepackage{natbib}

\usepackage{ucs}
\usepackage{upgreek}
\usepackage{longtable}

\usepackage{color}
\usepackage[normalem]{ulem}

\begin{document}
%


\title{Exploring GLIMPSE Bubble N107}
\subtitle{Multiwavelength Observations and Simulations}

\author{
  V.~Sidorin \inst{1} \and
  K.~A.~Douglas \inst{2,3,4} \and
  J.~Palouš \inst{1} \and
  R.~Wünsch \inst{1} \and
  S.~Ehlerová \inst{1}
}

\institute{
  Astronomical Institute of the Academy of Sciences of the Czech Republic, v.~v.~i.,
    Boční~II 1401/1a, 141~31, Praha~4, Czech Republic,
    \email{vojtech.sidorin@gmail.com} \and
  Department of Physics and Astronomy, University of Calgary,
    2500 University Dr NW, Calgary, Alberta, Canada, T2N~1N4 \and
  Dominion Radio Astrophysical Observatory,
    PO Box 248, Penticton, B.C.\ Canada, V2A~6J9 \and
  Okanagan College, 1000 KLO Road, Kelowna, B.C.\ Canada, V1Y 4X8
}

\date{Accepted 11/02/2014}


\abstract
  {
   Bubble N107 was discovered in the infrared emission of dust in the Galactic Plane observed by
   the Spitzer Space Telescope (GLIMPSE survey: $l \approx 51\fdg0$, $b \approx 0\fdg1$).
   The bubble represents an example of shell-like structures found all over the Milky Way Galaxy.
  }{
   We aim to analyse the atomic and molecular components of N107, as well as its radio continuum emission.
   With the help of numerical simulations, we aim to estimate the bubble age and other parameters
   which cannot be derived directly from observations.
  }{
   From the observations of the \ion{H}{i} (I-GALFA) and \element[][13]{CO} (GRS) lines
   we derive the bubble's kinematical distance and masses of the atomic and molecular components.
   With the algorithm DENDROFIND, we decompose molecular material into individual clumps.
   From the continuum observations at $1420\ \mathrm{MHz}$ (VGPS) and $327\ \mathrm{MHz}$ (WSRT),
   we derive the radio flux density and the spectral index.
   With the numerical code \emph{ring}, we simulate the evolution of stellar-blown bubbles similar to N107.
  }{
   The total \ion{H}{i} mass associated with N107 is $5.4 \times 10^3\ M_\odot$.
   The total mass of the molecular component (a mixture of cold gasses of H$_2$, CO, He and heavier elements)
   is $1.3 \times 10^5\ M_\odot$, from which $4.0 \times 10^4\ M_\odot$ is found along the bubble border.
   We identified 49 molecular clumps distributed along the bubble border, with the slope of the clump mass function of $-1.1$.
   The spectral index of $-0.30$ of a strong radio source located apparently within the bubble indicates nonthermal emission,
   hence part of the flux likely originates in a supernova remnant, not yet catalogued.
   The numerical simulations suggest N107 is likely less than $2.25\ \mathrm{Myr}$ old.
   Since first supernovae explode only after $3\ \mathrm{Myr}$ or later, no supernova remnant should be present within the bubble.
   It may be explained if there is a supernova remnant in the direction towards the bubble, however not associated with it.
  }{
  }





\maketitle


%
%

\section{Introduction}


Bubble N107 (SIMBAD: \object{CPA2006 N107}) is one of the largest bubbles in the catalogue by \citet{2006apj...649..759c}.
This catalogue is based on the infrared observations made with the Spitzer Space Telescope,
(GLIMPSE survey, \citealp{2003pasp..115..953b}) and contains more than 300 bubbles
found in the interstellar medium (ISM) near the Galactic Plane.
The catalogue was later expanded by \citet{2012mnras.424.2442s} to more than 5000 bubbles,
which were identified by a community of over $35\,000$ volunteers.
These bubbles, with typical radii less than $10\ \mathrm{pc}$,
are examples of shell-like structures -- features found in the ISM all over our Galaxy.
Other examples are \ion{H}{i} shells in catalogues by \citet{1979apj...229..533h,1984apjs...55..585h,2005a&a...437..101e},
with sizes up to $3500\ \mathrm{pc}$; or
dusty ``loops'' discovered in the IRAS full-sky survey by \citet{2004a&a...418..131k}
and \citet{2007a&a...463.1227k}.

The varying sizes and masses of these structures suggest different
mechanisms responsible for their creation.
These mechanisms, which do not have to act solely, include:

\begin{enumerate}
  \item feedback from massive (OB) stars: radiation, winds and supernovae (see refs.\ below);
  \item infall of high-velocity clouds into the Galactic disc
        (e.g. \citealp{1984apjs...55..585h,1988ara&a..26..145t,1996a&a...313..478e});
  \item energy and mass inserted by gamma-ray bursts
        (e.g. \citealp{1998apj...501l.163e,1998apj...503l..35l,1999a&a...350..457e}); and
  \item turbulence
        (\citealp{2005apj...630..238d}).
\end{enumerate}

\noindent
Bubble N107, as most of the GLIMPSE bubbles,
is likely a result of stellar feedback \citep{2006apj...649..759c,2010a&a...523a...6d},
which may follow three different models:

\begin{enumerate}
  \item An expanding \ion{H}{ii} region, assuming radiative heating and ionisation of hydrogen
        by UV photons from a massive star \citep{1954ban....12..187k,1955apj...121....6O,1978ppim.book.....s}.
        The pressure inside the \ion{H}{ii} region is higher compared to the ambient neutral medium,
        so the \ion{H}{ii} region expands and collects the neutral material in an expanding shell.
  \item A wind-blown bubble, supposing that energy is released via stellar winds \citep{1975apj...200l.107c,1977apj...218..377w}.
        Such a bubble, formed around a massive star, can be divided into several layers:
        an innermost layer, where the wind is freely expanding up to a layer of shocked stellar wind,
        where its kinetic energy is thermalised.  This layer of the shocked wind is followed by
        a layer of swept up ISM, which in later stages of evolution is mainly atomic and molecular.
  \item A supernova explosion, supposing an abrupt energy input into the ISM, producing an expanding
        shell/remnant \citep[][and citations therein]{1974apj...188..501c}.
\end{enumerate}

\noindent
These different models of stellar feedback were also discussed by
\citet{1988ara&a..26..145t,1996a&a...313..478e} and \citet{1999a&a...350..457e}.

The shell-like structures can trigger star formation, since they consist of a layer of cold
and dense material.
In the so-called collect-and-collapse scenario
\citep{1977apj...214..725e,1994apj...427..384e,1994mnras.268..291w},
the shell fragments and creates a new generation of stars.
Radiation driven implosion \citep{1990a&a...233..190d,1994a&a...289..559l}
considers the compression of preexisting condensations (globules) by the pressure of the ionised gas.
A brief review of these and other triggering mechanisms is given in a study of GLIMPSE bubbles
by \citet{2010a&a...523a...6d}.

\begin{figure*}
  \includegraphics[width=\hsize]{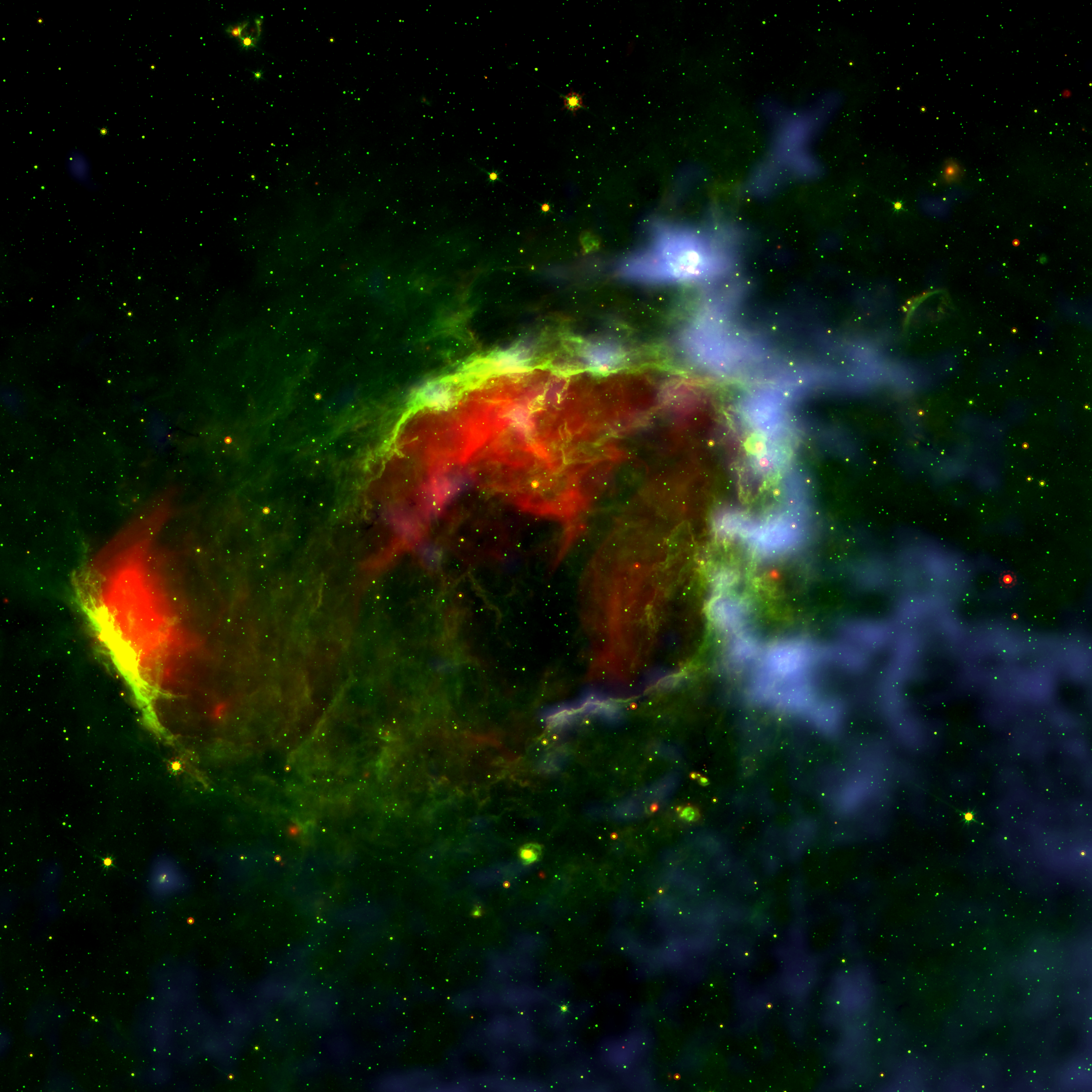}
  \caption{Bubble N107 -- false colour multiwavelength image.  Red corresponds to
    $24\ \mathrm{\upmu m}$ continuum, green to $8\ \mathrm{\upmu m}$ continuum and
    blue to \element[][13]{CO} ($J=1\mathrm{-}0$) line integrated over the radial velocities of
    $38.5$ to $47.6\ \mathrm{km/s}$.
    Note that the $24\ \mathrm{\upmu m}$ continuum (red) is dominated
    by the emission of very small dust grains heated by a nearby massive star.
    The $8\ \mathrm{\upmu m}$ continuum (green) is dominated by the emission of PAHs --
    large molecules found within molecular clouds.
    The emission of PAHs is tracing the photodissociation regions (PDRs) -- edges of these
    clouds illuminated by UV radiation from a massive star.
    The CO line (blue) traces the cold gas of molecular clouds.
    The distinct red-green structure seen in the lower-left in the direction of the bubble's opening
    is likely not related to the bubble, since the radial velocity of the associated molecular
    material ($\approx 60\ \mathrm{km/s}$) differs significantly from the radial velocity of the N107 complex
    ($\approx 43\ \mathrm{km/s}$).
    }
  \label{fig_n107_multi_co}
\end{figure*}

In this paper, we present a multiwavelength study of one of the largest GLIMPSE bubbles: N107,
which lies in the central plane of the Galactic disc ($l \approx 51\fdg0$, $b \approx 0\fdg1$).
A multiwavelength, colour picture of N107 is shown in fig.~\ref{fig_n107_multi_co}.

The paper is organised in the following way:
In section~\ref{sec_observation}, we complement the Spitzer infrared observations
with \ion{H}{i} line observations from the I-GALFA survey \citep{2010hia....15..788k},
\element[][13]{CO} ($J=1\mathrm{-}0$) line observations from the GRS survey \citep{2006apjs..163..145j}
and radio continuum observations at $1420$ and $327\ \mathrm{MHz}$ from the
VGPS \citep{2006aj....132.1158s} and WSRT \citep{1996apjs..107..239t} surveys.
From the line observations, we derive the bubble's LSR\footnote{LSR = local standard of rest} radial velocity (RV),
kinematical distance and masses of the molecular and atomic components.
From the radio continuum observations, we derive the flux densities and corresponding spectral indices
of two radio sources lying apparently inside the bubble.
In section~\ref{sec_analysis_of_molecular_clumps}, we use a recently published algorithm DENDROFIND \citep{2012a&a...539a.116w} to
decompose the molecular gas associated with the bubble into individual clumps
and derive the slope of the clump mass function.
In section~\ref{sec_simulations}, we use the numerical code \emph{ring} \citep{1990iaus..144p.101p,1996a&a...313..478e} to
simulate the bubble's evolution in order to estimate its age, energy input and the size and mass of
the molecular cloud from which the bubble evolved.
Finally, a discussion and conclusions are presented in sections~\ref{sec_discussion} and \ref{sec_conclusion}.

%
%

\section{Observation of Bubble N107\label{sec_observation}}

\subsection{Data Sources\label{sec_data_sources}}

\emph{GLIMPSE} (Galactic Legacy Infrared Mid-Plane Survey Extraordinaire, \citealp{2003pasp..115..953b})
and \emph{MIPSGAL} \citep{2009pasp..121...76c} are two complementary infrared surveys mapping
the Galactic disc with the Spitzer Space Telescope.  GLIMPSE used instrument IRAC (Infrared Array
Camera) which has four channels centred around $3.6$, $4.5$, $5.8$ and $8.0\ \mathrm{\upmu m}$, for
which the point response function FWHM is between $1\farcs7$ and $2\farcs0$.
MIPSGAL used the instrument MIPS (Multiband Imaging Photometer for Spitzer) with a broadband channel
centred around $24\ \mathrm{\upmu m}$, for which the point spread function FWHM is $6\arcsec$.

\emph{I-GALFA} (Inner-Galaxy ALFA Low-Latitude \ion{H}{i} Survey, \citealp{2010hia....15..788k})
is part of a large project called GALFA-\ion{H}{i} \citep{2011apjs..194...20p},
which plans to map the entire sky observable by the Arecibo Observatory in
the \ion{H}{i} line.
The FWHM beam size of the Arecibo telescope at $1420\ \mathrm{MHz}$ is $3\farcm35$
with the GALFA-\ion{H}{i} radial velocity channel width of $0.184\ \mathrm{km/s}$.
The data are provided in FITS cubes of the brightness temperature $T_\mathrm{b}$.
The single-channel standard deviation of the gaussian noise $\sigma_\mathrm{noise}$
is about $0.25\ \mathrm{K}$.

The \emph{GRS} (Galactic Ring Survey, \citealp{2006apjs..163..145j})
is a \element[][13]{CO} ($J=1\mathrm{-}0$) survey focused on the Galactic molecular ring.
The survey used the FCRAO 14-m telescope, with a
FWHM beam size of $46\arcsec$ and a radial velocity channel width of $0.21\ \mathrm{km/s}$.
The data are provided in FITS cubes of the antenna temperature $T_\mathrm{A}^*$,
with a typical noise $\sigma_\mathrm{noise}^* \approx 0.13\ \mathrm{K}$.
We converted the data to the main-beam brightness temperature:
$T_\mathrm{b} = T_\mathrm{A}^* / \eta_\mathrm{MB}$ and
$\sigma_\mathrm{noise} = \sigma_\mathrm{noise}^* / \eta_\mathrm{MB}$,
where $\eta_\mathrm{MB} = 0.48$ is the main beam efficiency (value suggested by
\citealp{2006apjs..163..145j}).

\emph{VGPS} (VLA Galactic Plane Survey, \citealp{2006aj....132.1158s})
is a $1420\ \mathrm{MHz}$ continuum and \ion{H}{i} line survey of the Galactic disc
based on the interferometric observations done with the VLA (Karl G.\ Jansky Very Large Array).
We use only the continuum observations since for the \ion{H}{i}, data from I-GALFA are available
with significantly lower noise.
The FWHM beam size is $1\arcmin$ and the gaussian noise is $0.3\ \mathrm{K}$, for the continuum observations.

\emph{WSRT} (Westerbork Synthesis Radio Telescope survey, \citealp{1996apjs..107..239t})
is a $327\ \mathrm{MHz}$ continuum survey of the Galactic plane utilising the WSRT interferometer
in the Netherlands.  The FWHM resolution is $1\arcmin \times (1\arcmin / \sin\delta)$ in the right ascension
and declination ($\delta$) direction, respectively.
This yields for N107 ($\delta \approx 16\degr$) the FWHM resolution of $1\arcmin \times 3\farcm6$.
The median gaussian noise is $2.5\ \mathrm{mJy/beam}$.

The \emph{UKIDSS} (UKIRT Infrared Deep Sky Survey) project is defined in \citet{2007MNRAS.379.1599L}.
UKIDSS uses the UKIRT (United Kingdom Infrared Telescope) Wide Field Camera (WFCAM; \citealp{2007A&A...467..777C})
and a photometric system described in \citet{2006MNRAS.367..454H}.
The pipeline processing and science archive are described in \citet{2004SPIE.5493..411I} and \citet{2008MNRAS.384..637H}.
We used data from the 7th data release.

\subsection{Dust Component\label{sec_dust_component}}


The bubble's outline is most prominent in the $8\ \mathrm{\upmu m}$ maps from the GLIMPSE survey,
where it was discovered by \citet{2006apj...649..759c}, who derived its basic properties:
mean position of $l \approx 51\fdg0$, $b \approx 0\fdg1$; mean radius of $11\farcm4$ and a mean thickness
of $2\farcm3$.  Most emission at $8\ \mathrm{\upmu m}$ is coming from the bubble edges,
forming a ring-like structure.
The $8\ \mathrm{\upmu m}$ channel is dominated by the emission of polycyclic aromatic hydrocarbons (PAHs)
(\citealp{2003ara&a..41..241d}; \citealp[][fig.~1]{2006aj....131.1479r}; \citealp{2013arep...57..573p}).
PAHs, which are destroyed in the ionised regions,
are tracing the photodissociation regions (PDRs), edges of molecular clouds
illuminated by the UV radiation from a massive star.  The UV radiation pervades
the PDRs and excites the PAHs herein.

Counterpart emission at $24\ \mathrm{\upmu m}$ was observed in MIPSGAL \citep{2009pasp..121...76c}.
Contrary to the $8\ \mathrm{\upmu m}$ continuum, emission at
$24\ \mathrm{\upmu m}$ is filling part of the bubble's interior.
The $24\ \mathrm{\upmu m}$ continuum is dominated by the emission of
very small grains -- dust grains larger than PAHs --
heated by a nearby massive star \citep{2013arep...57..573p}.

A dark hole is present in the southwestern part of the bubble's interior.  Almost no emission from dust,
and also no emission of \element[][13]{CO} or \ion{H}{i} (see below), is observed there.
Fig.~\ref{fig_n107_multi_co} shows a multiwavelength, colour picture of N107
composed of the $8\ \mathrm{\upmu m}$ and $24\ \mathrm{\upmu m}$ continuum emission
and the \element[][13]{CO} line emission integrated over the radial velocities of $38.5$ to $47.6\ \mathrm{km/s}$.

\subsection{Atomic Component\label{sec_atomic_component}}

\begin{figure*}
  
  \includegraphics[width=.24\hsize]{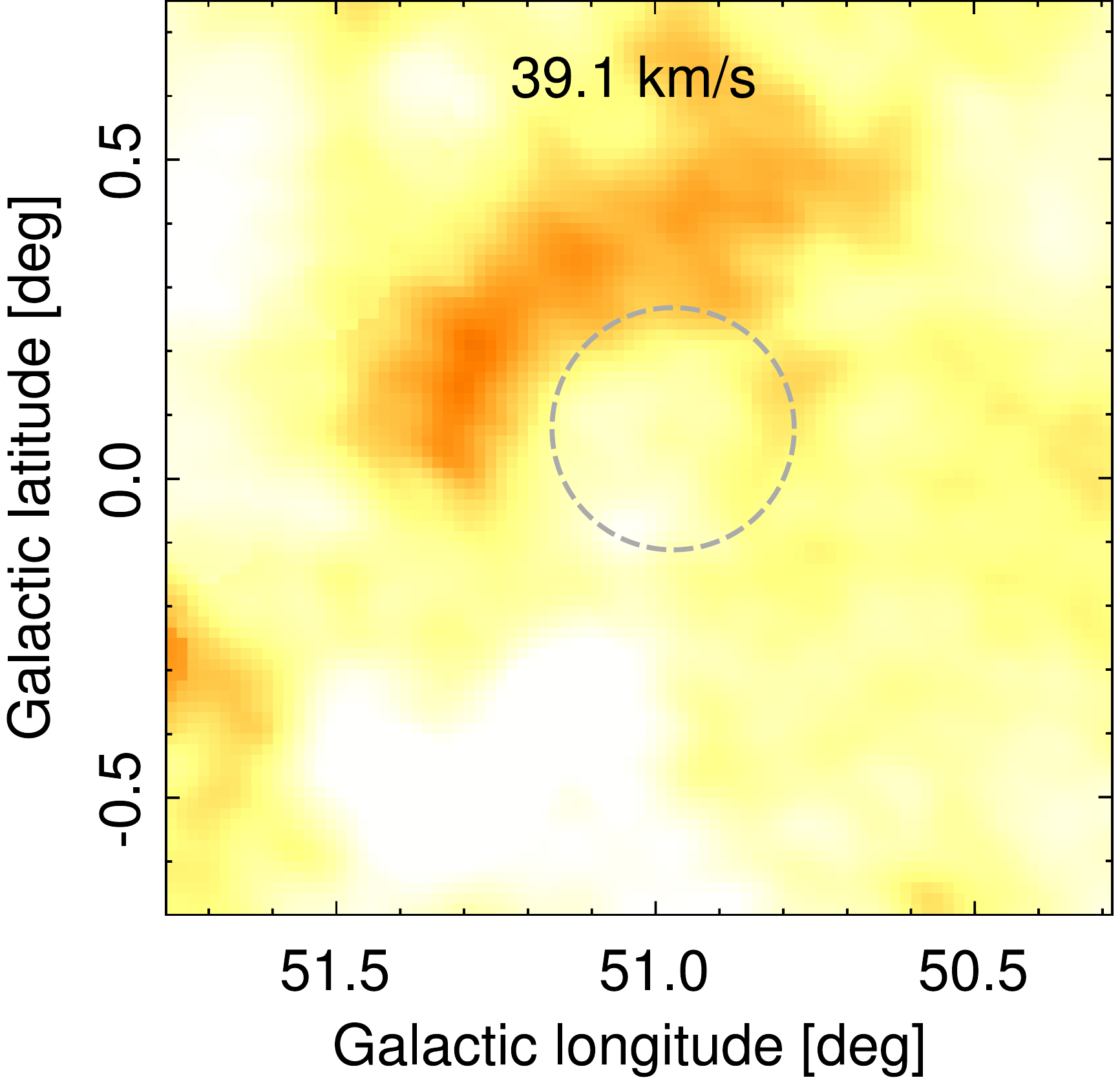}\hfill
  \includegraphics[width=.24\hsize]{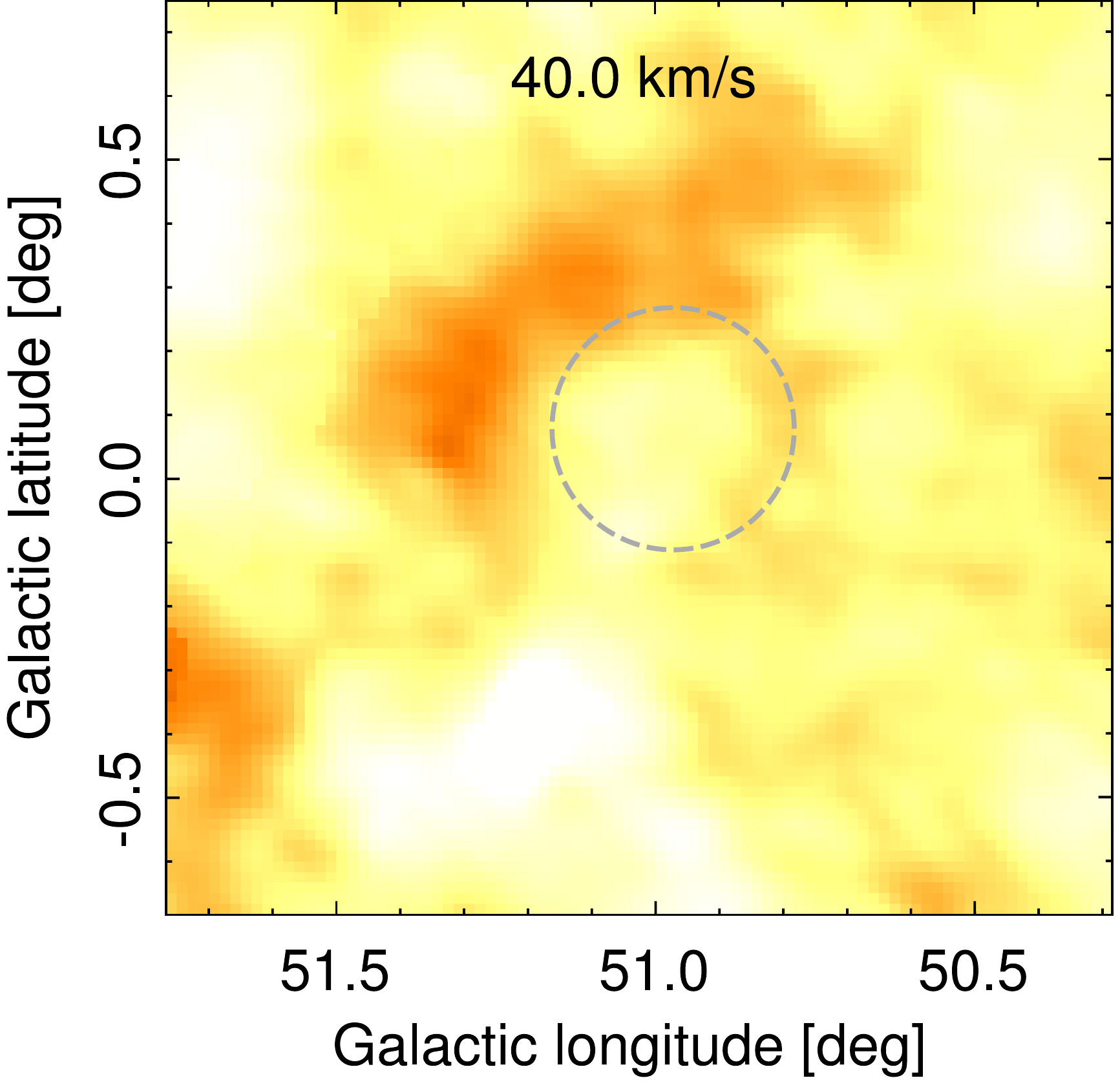}\hfill
  \includegraphics[width=.24\hsize]{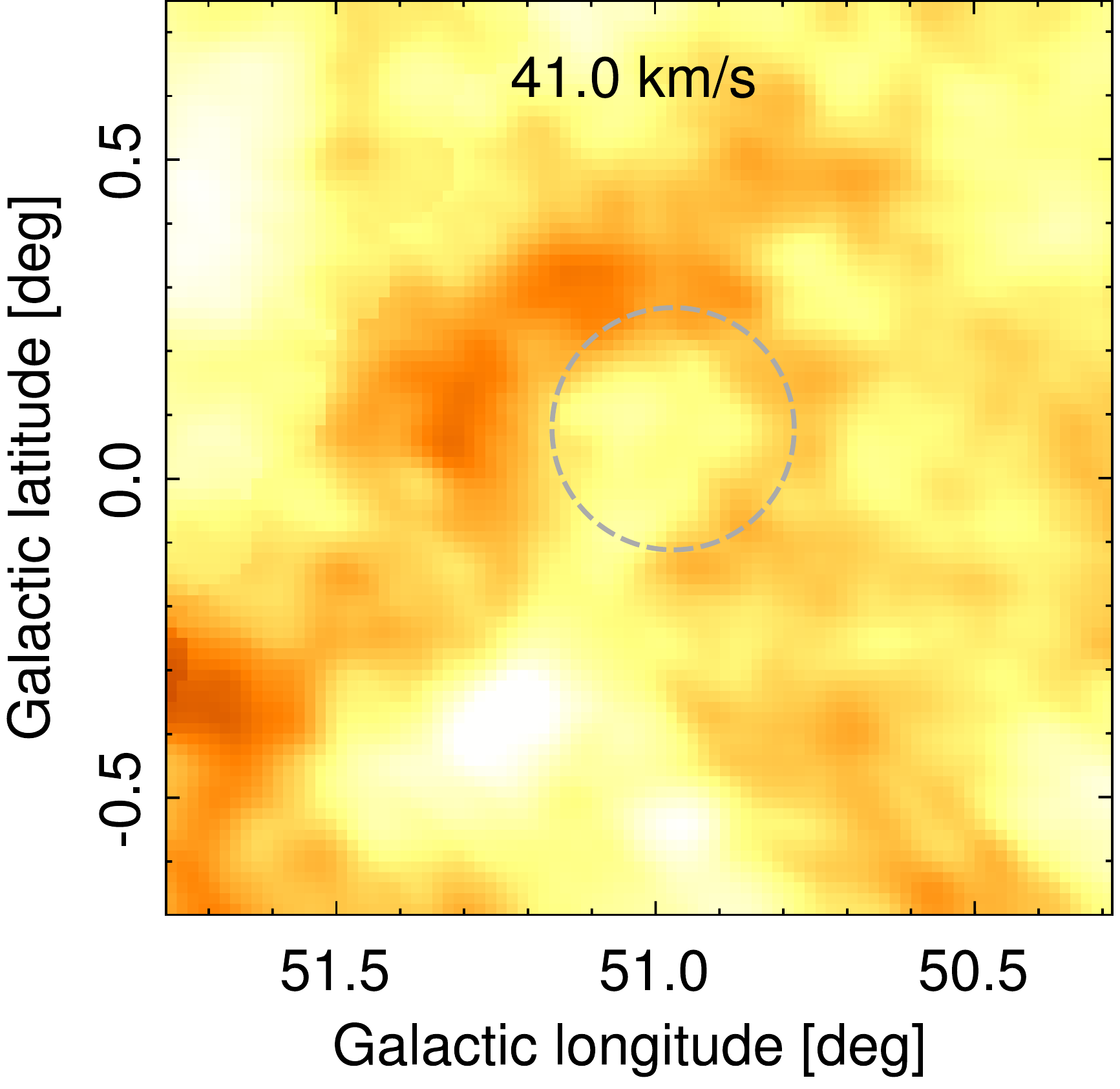}\hfill
  \includegraphics[width=.24\hsize]{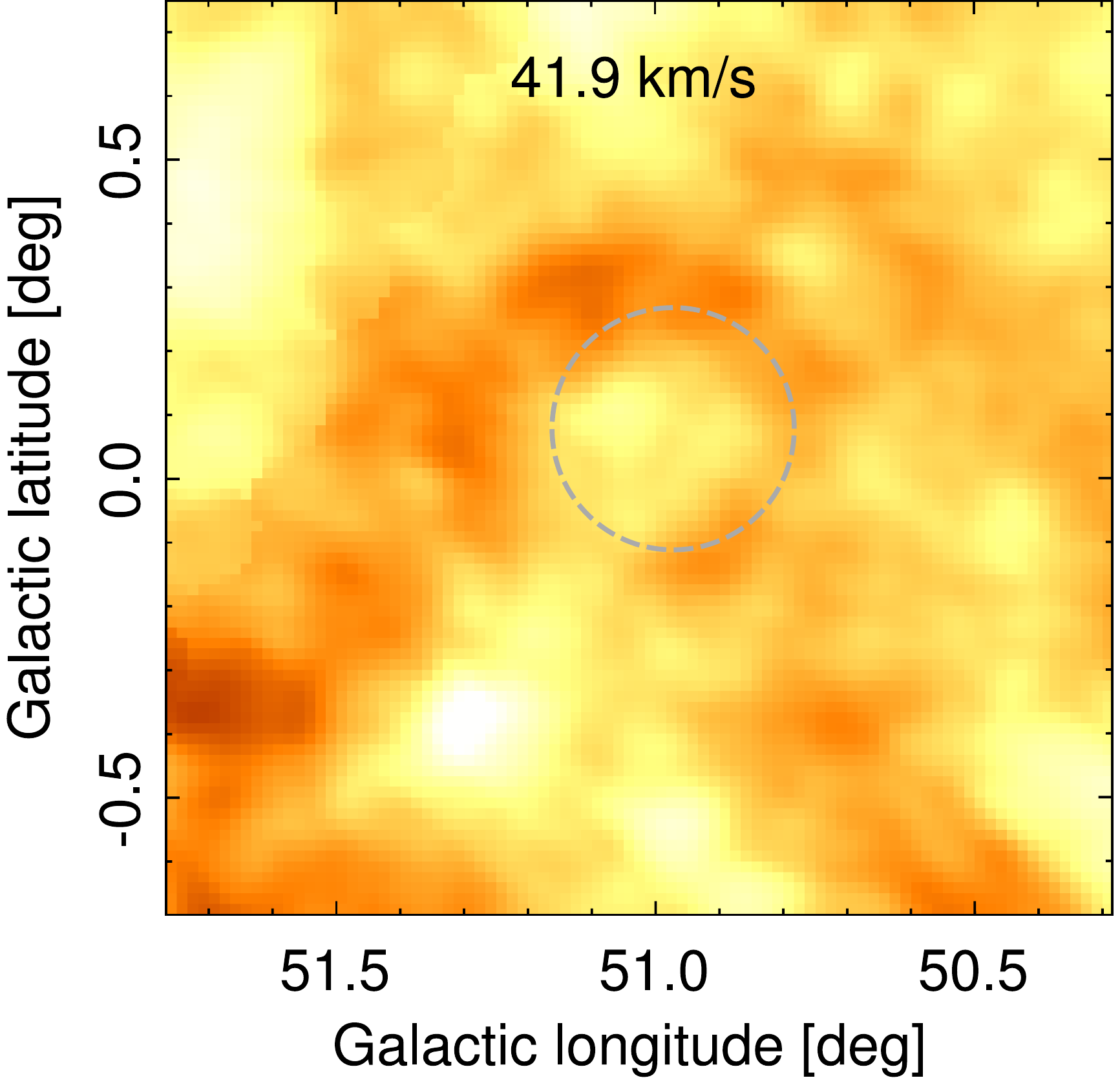}\hfill
  
  \medskip\noindent
  
  \includegraphics[width=.24\hsize]{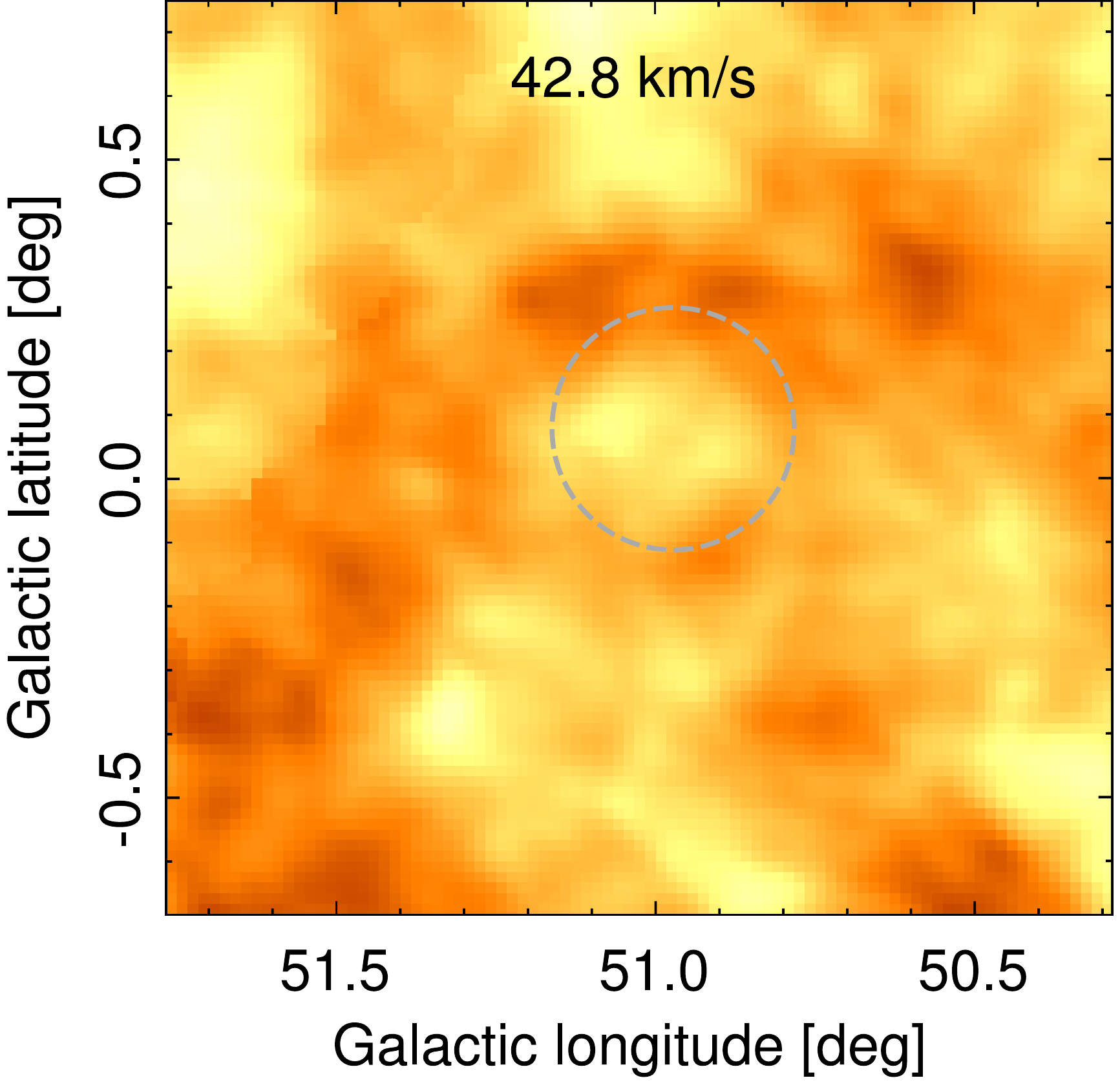}\hfill
  \includegraphics[width=.24\hsize]{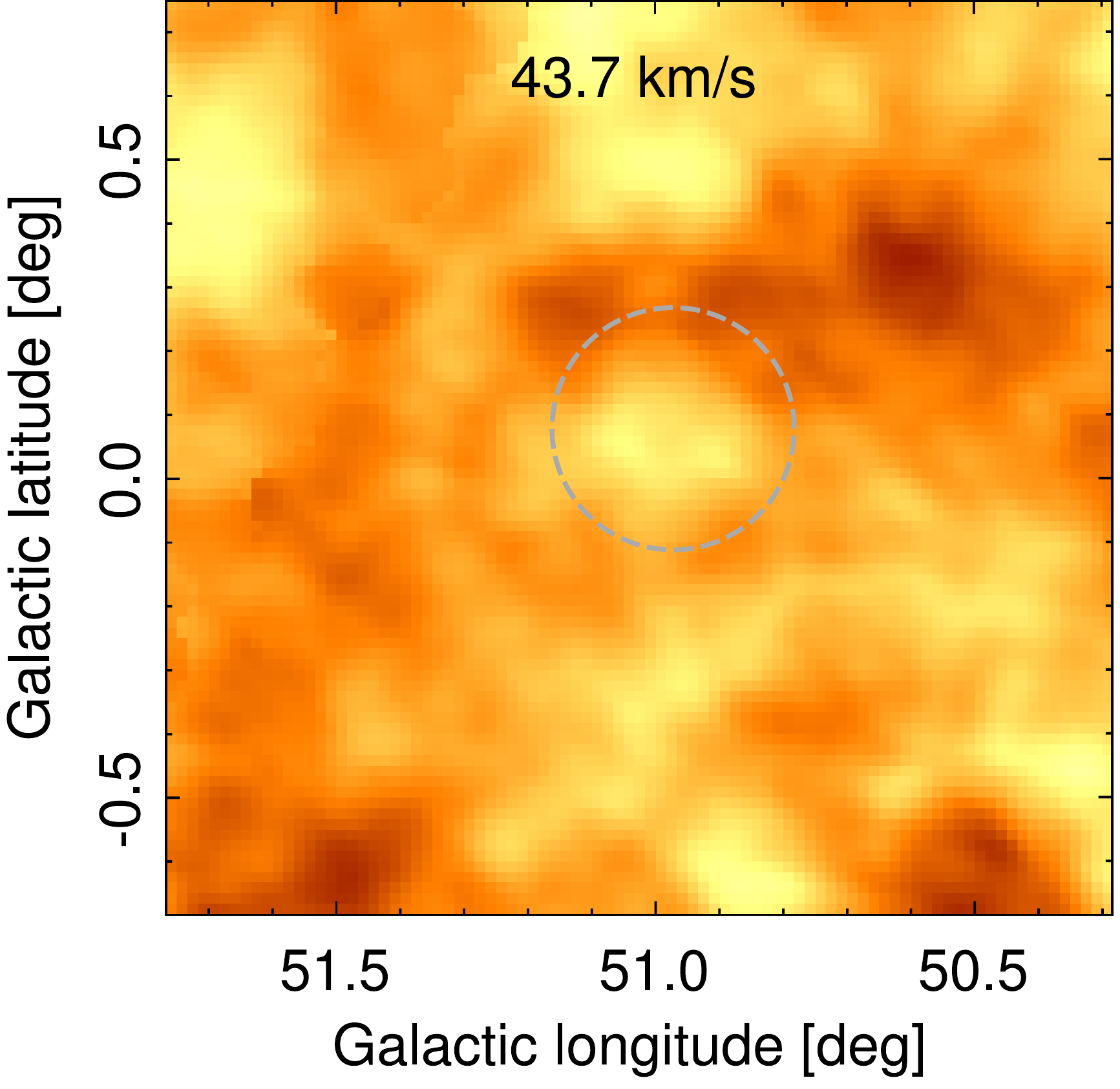}\hfill
  \includegraphics[width=.24\hsize]{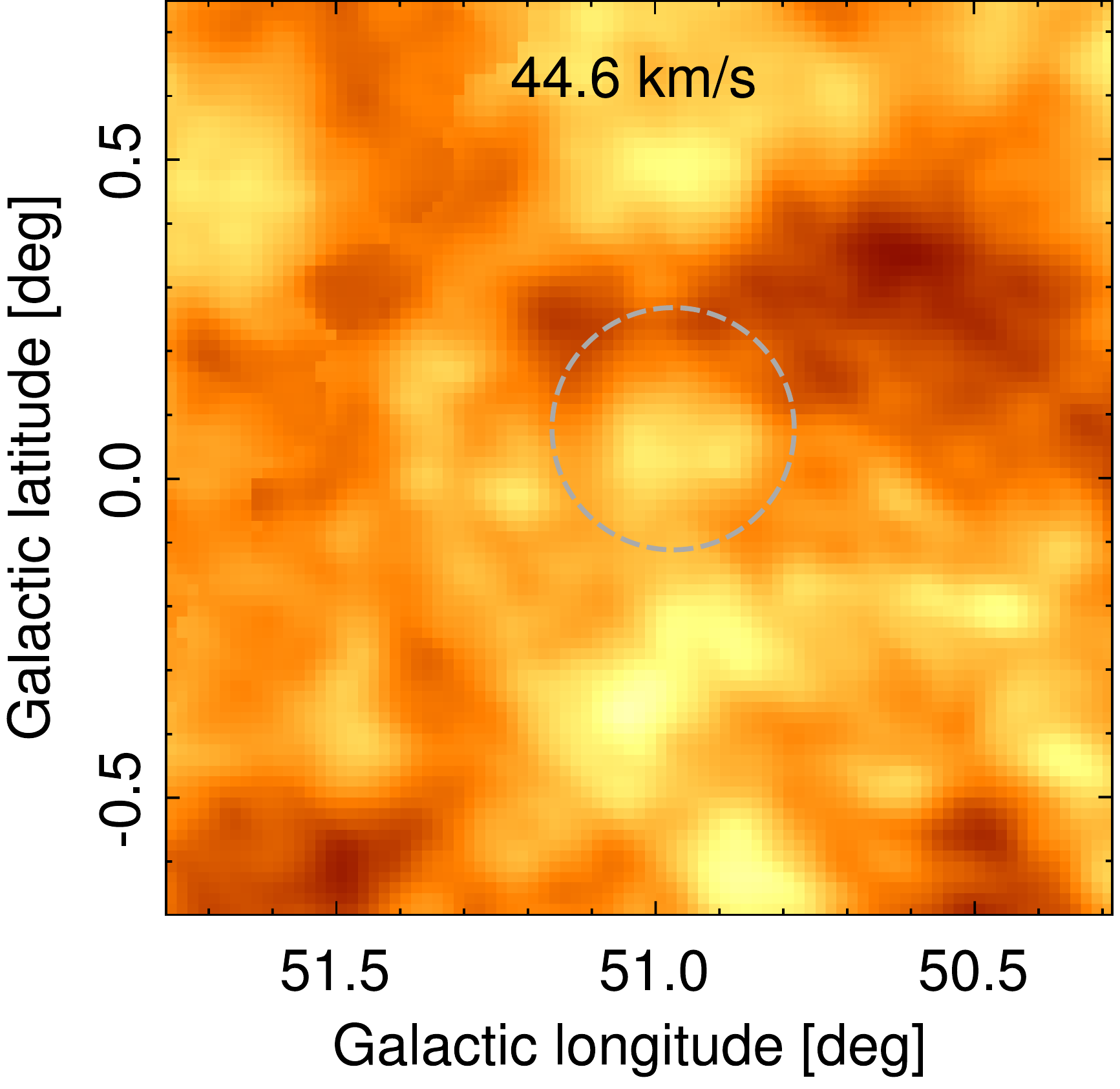}\hfill
  \includegraphics[width=.24\hsize]{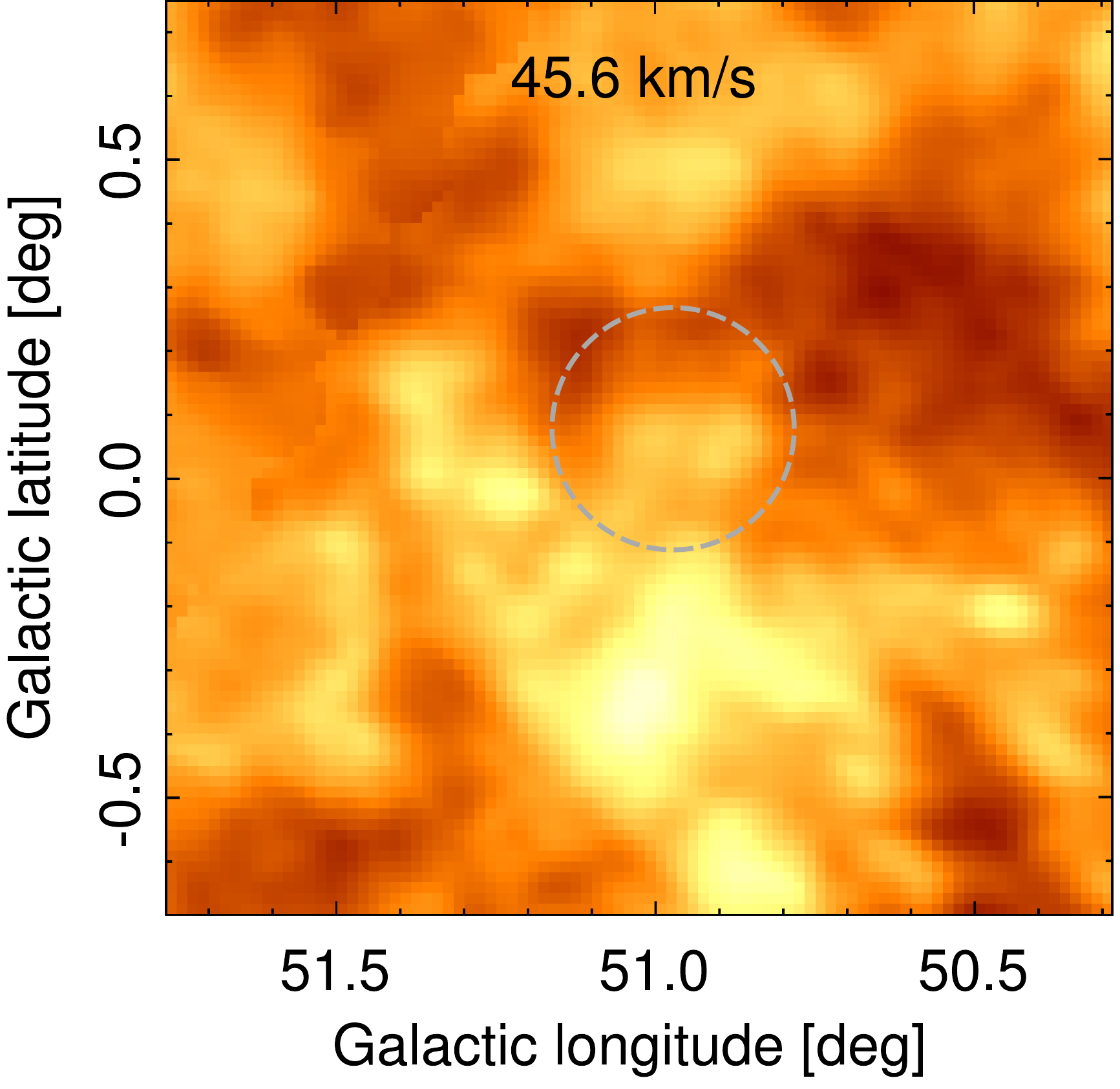}\hfill
  
  \medskip\noindent
  
  \includegraphics[width=\hsize]{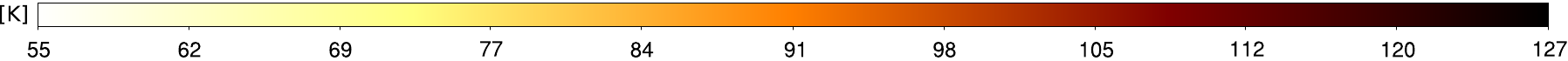}
  
  \caption{Brightness temperature maps of the \ion{H}{i} line in several velocity
    channels.  An outline of bubble N107 as given in the catalogue of
    \citet{2006apj...649..759c} is marked with the dashed grey circle.
    Especially the channels $41.0$ and $41.9\ \mathrm{km/s}$ show morphology
    similar to that of the $8\ \mathrm{\upmu m}$ emission.}
  \label{fig_igalfa}
\end{figure*}

\begin{figure}
  \includegraphics[width=\hsize]{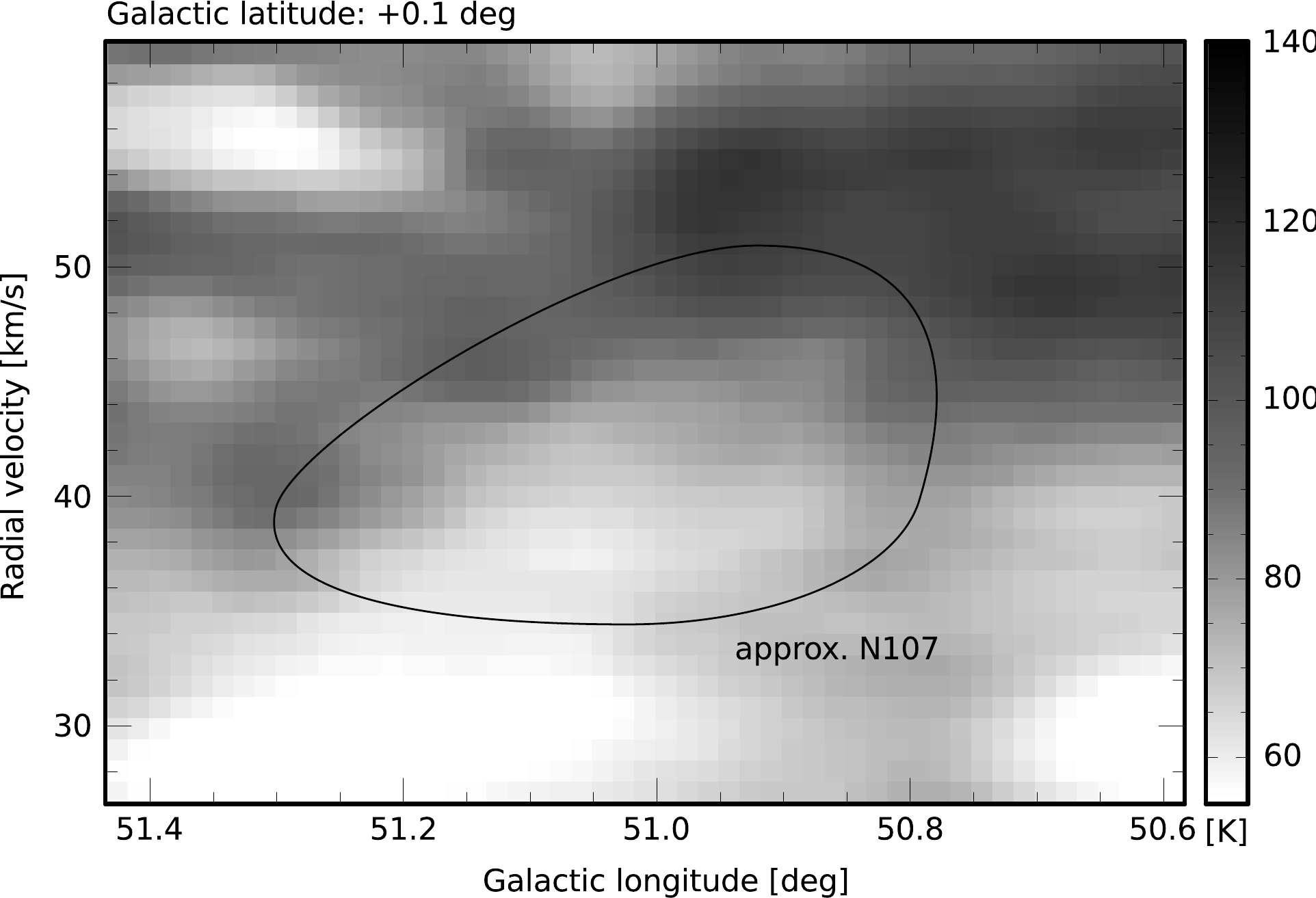}
  \caption{Brightness temperature of the \ion{H}{i} line in an $l$--$v$ diagram: a cut through $b = 0\fdg1$.
    The black curve marks the approximate extent of the \ion{H}{i} shell associated with N107.}
  \label{fig_igalfa_lv}
\end{figure}

\begin{figure*}
  \includegraphics[width=.49\hsize]{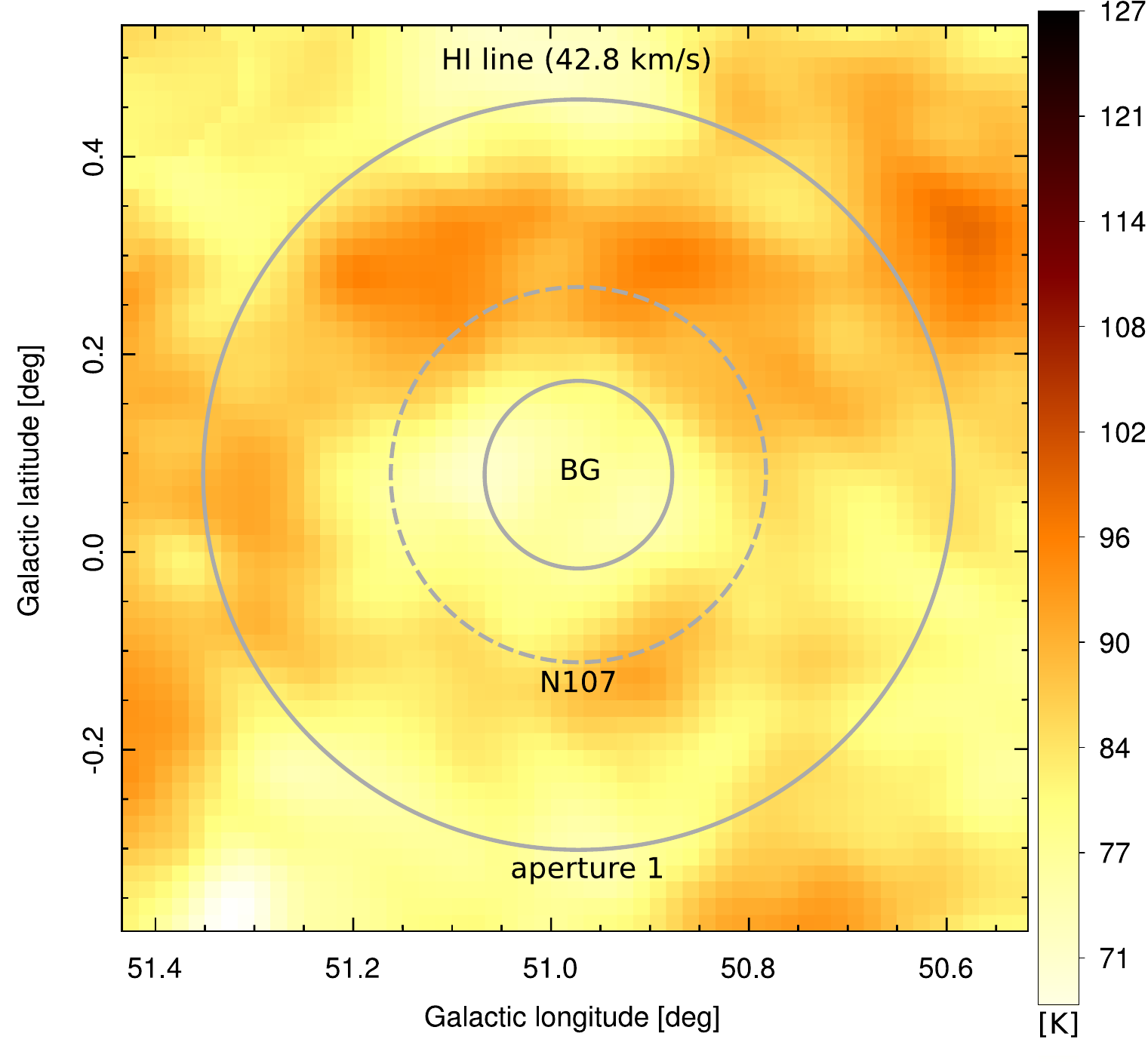}\hfill
  \includegraphics[width=.49\hsize]{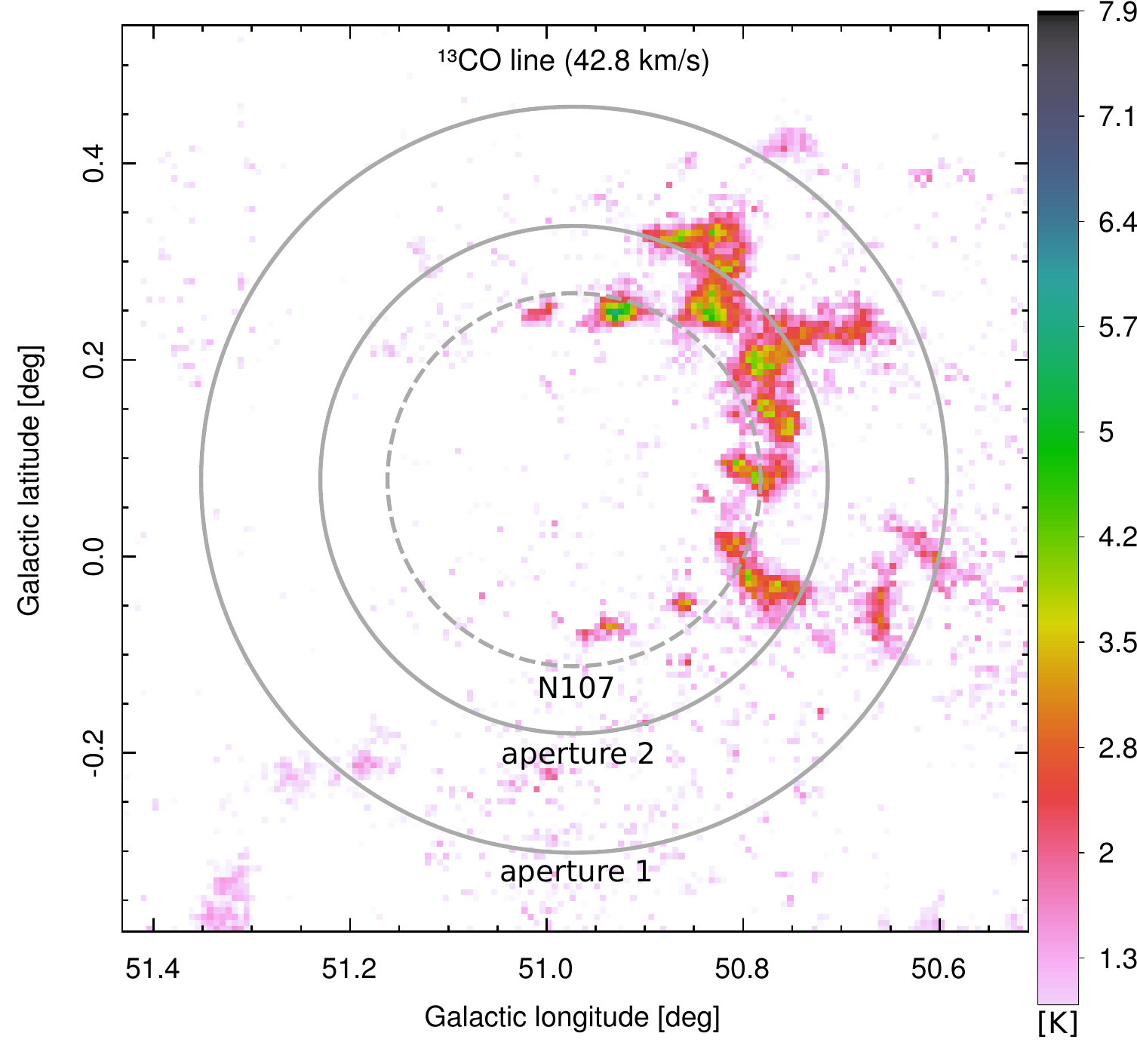}
  \caption{Apertures for measuring the mass associated with bubble N107.
    The pixels lying within the apertures contribute to the measured masses.
    The dashed circles mark bubble N107 as was identified in the $8\ \mathrm{\upmu m}$ emission
    by \citet{2006apj...649..759c}.
    \textbf{Left:} \ion{H}{i} line emission at $42.8\ \mathrm{km/s}$.  The innermost circle (BG) is the
    region we used to derive the background emission.
    \textbf{Right:} \element[][13]{CO} line emission at $42.8\ \mathrm{km/s}$.
    We use two apertures for measuring the molecular mass.  Aperture 1 is same as that for measuring
    the \ion{H}{i} mass, while aperture 2 is smaller and covers only the immediate vicinity of the bubble.
    }
  \label{fig_apertures}
\end{figure*}

We searched the \ion{H}{i} data cubes by eye for features possibly associated with bubble N107.
An \ion{H}{i} shell, morphologically similar to N107, is located around the central LSR radial
velocity of $43\ \mathrm{km/s}$.
Note that an incomplete ring of CO clumps is present at similar radial velocities (see sec.~\ref{sec_molecular_component}).
The \ion{H}{i} emission is located along the bubble edges, protruding outside farther than the CO emission,
forming an atomic envelope of the whole structure.
Maps of the \ion{H}{i} brightness temperature in several velocity channels are shown in fig.~\ref{fig_igalfa}.

Fig.~\ref{fig_igalfa_lv} shows a map of the galactic longitude versus radial velocity ($l$--$v$ diagram).
The front wall is very faint, located at a radial velocity of $\approx 34\ \mathrm{km/s}$.
The back wall is more distinct, located at a radial velocity of $\approx 50\ \mathrm{km/s}$.
Between these radial velocities, a hole is apparent inside the bubble.
The relative expansion velocity of the front versus back wall is $\approx 16\ \mathrm{km/s}$.

In order to measure the \ion{H}{i} mass associated with the bubble,
we define the following 3D aperture (aperture 1, see also fig.~\ref{fig_apertures}, left panel):
radius of two times that of N107, $r_\mathrm{ap1} = 2 \cdot r_\mathrm{N107} = 22\farcm8$,
and the radial velocity channels from $32.7$ to $52.0\ \mathrm{km/s}$ inclusive.
All the pixels lying within this aperture contribute to the measured \ion{H}{i} mass.

To derive the mass of the neutral hydrogen, we use the optically thin limit of \ion{H}{i} line radiative transfer.
Then, the \ion{H}{i} column density $N_\ion{H}{i}$ can be computed from the formula \citep{1996tra..book.....r}:

\begin{equation}
  \frac{N_\ion{H}{i}}{\mathrm{cm}^{-2}} = 1.8 \times 10^{18} \int \frac{T_\mathrm{b}}{\mathrm{K}} \frac{\mathrm{d}v}{\mathrm{km/s}},
\end{equation}

\noindent
where $T_\mathrm{b}$ is the observed brightness temperature of the \ion{H}{i} gas and,
in our case of an I-GALFA data cube, the integral takes form of a sum over a set of pixels with
$\mathrm{d}v = 0.184\ \mathrm{km/s}$.

The Galactic \ion{H}{i} emission has a strong background component.
To estimate it, we assume that the bubble's interior is evacuated,
so the \ion{H}{i} emission observed apparently near the bubble centre
is due to this \ion{H}{i} background, not associated with the bubble.
We defined another 3D aperture with the radius half that of the N107
and the radial velocity channels from $39.1$ to $43.7\ \mathrm{km/s}$ (see fig.~\ref{fig_apertures}).
This aperture encloses a volume in the central part of the bubble.
The mean $T_\mathrm{b}$ within this region is $72\ \mathrm{K}$,
which we adopt as the \ion{H}{i} background component
and subtract it before computing the mass.

Furthermore, in the process of mass derivation, we need to know the \textbf{distance to N107}.
Considering also the \element[][13]{CO} emission (see below),
we adopt the central radial velocity of $42.8\ \mathrm{km/s}$.
The Galactic rotation model of \citet{1993a&a...275...67b} gives for that radial velocity two kinematical distances:
$3.6$ and $7.1\ \mathrm{kpc}$.
We adopt the near one of $3.6\ \mathrm{kpc}$,
which yields the bubble mean radius of $11.9\ \mathrm{pc}$ ($11.4\arcmin$).
The near distance is favoured by the estimated virial masses of the molecular clumps found along the bubble borders.
For the far distance of $7.1\ \mathrm{kpc}$, we got 22 of 49 clumps with supervirial masses,
while for the near distance of $3.6\ \mathrm{kpc}$, we got only 2 clumps with virial masses above 1
(clumps 7 and 21 in tab.~\ref{tab_clumps_full}).
Note that \citet{2006apj...649..759c} suggest distances of $4.7$/$6.0\ \mathrm{kpc}$.
They use the same rotation model as we do, however with a less reliable determination of the radial velocity
obtained from the observation of coincident \ion{H}{ii} regions.
They advocate also the near kinematical distance.

To estimate the uncertainty of the measured mass, we use the standard Taylor formula:
\begin{equation}
    \label{eq_uncertainty_propagation}
    \sigma_f = \sqrt{\sum_{i}^{}\left(\frac{\partial f}{\partial x_i}\right)^2 \sigma_{x_i}^2},
\end{equation}

\noindent
where $f$ is the measured value (in our case the \ion{H}{i} mass),
$\sigma_f$ is its uncertainty, which is derived from uncertainties $\sigma_{x_i}$
of variables $x_i$ on which $f$ depends (in our case e.g. the distance to N107).

The resulting \ion{H}{i} mass associated with N107,
i.e. the total \ion{H}{i} mass within aperture 1 after subtracting the background,
is $5.4 \times 10^{3}\ M_\odot$.
The major sources of uncertainty are the distance to N107 and the background emission.
Assuming the distance uncertainty of $10\ \%$ ($0.36\ \mathrm{kpc}$)
and the background uncertainty of $1\ \mathrm{K}$,
the total \ion{H}{i} mass uncertainty is about $20\ \%$, i.e. $1 \times 10^{3}\ M_\odot$.

\subsection{Molecular Component\label{sec_molecular_component}}

\begin{figure*}
  \includegraphics[width=.24\hsize]{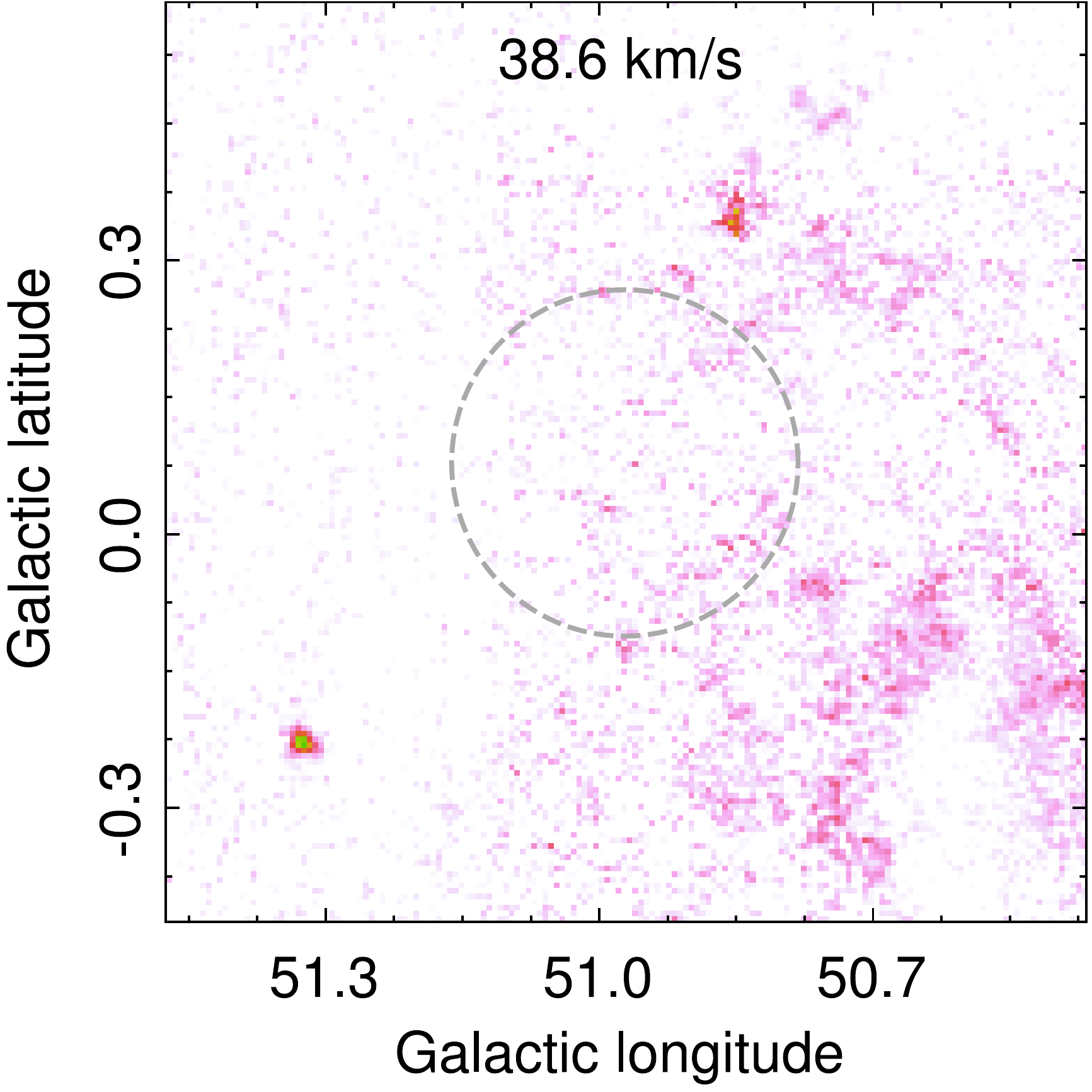}\hfill
  \includegraphics[width=.24\hsize]{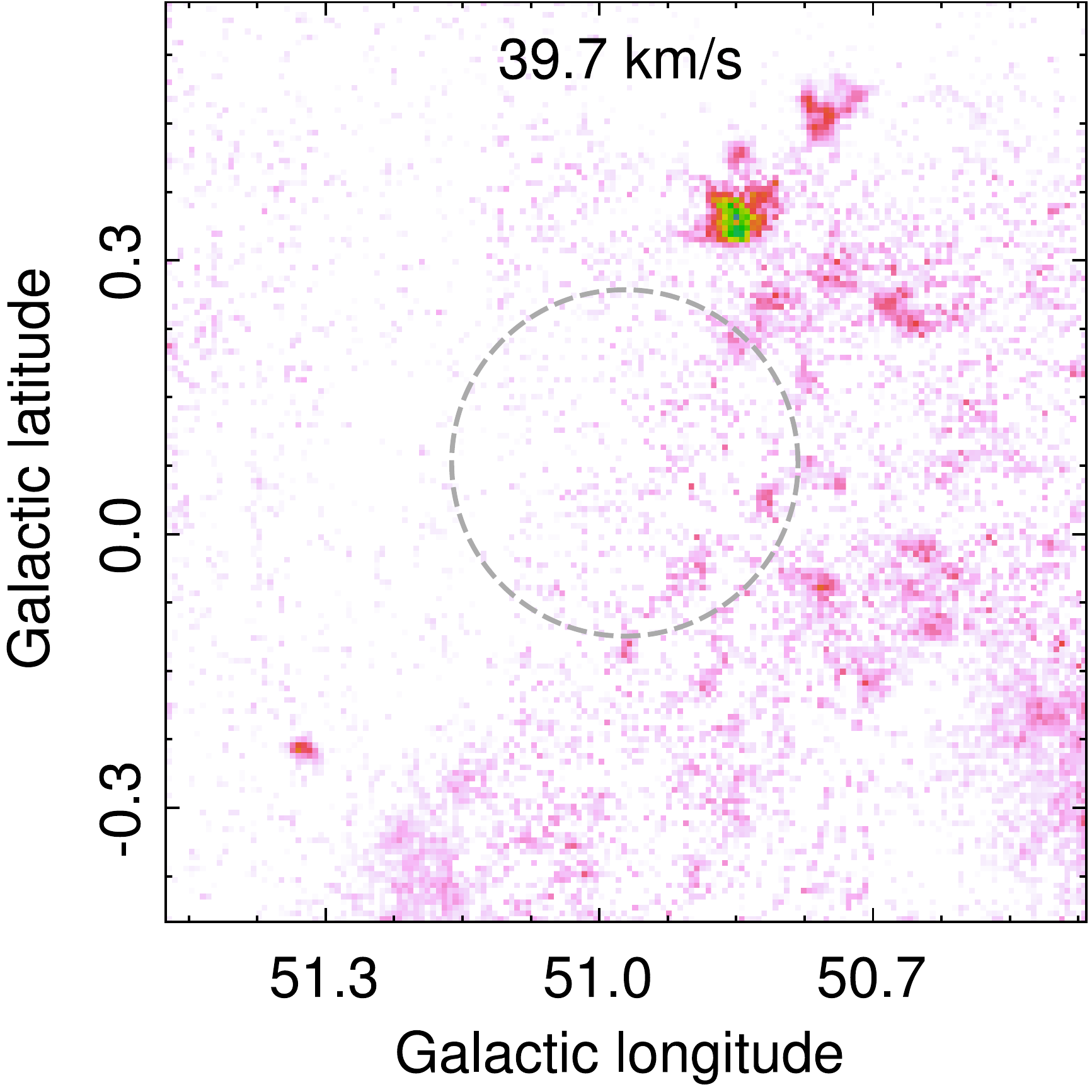}\hfill
  \includegraphics[width=.24\hsize]{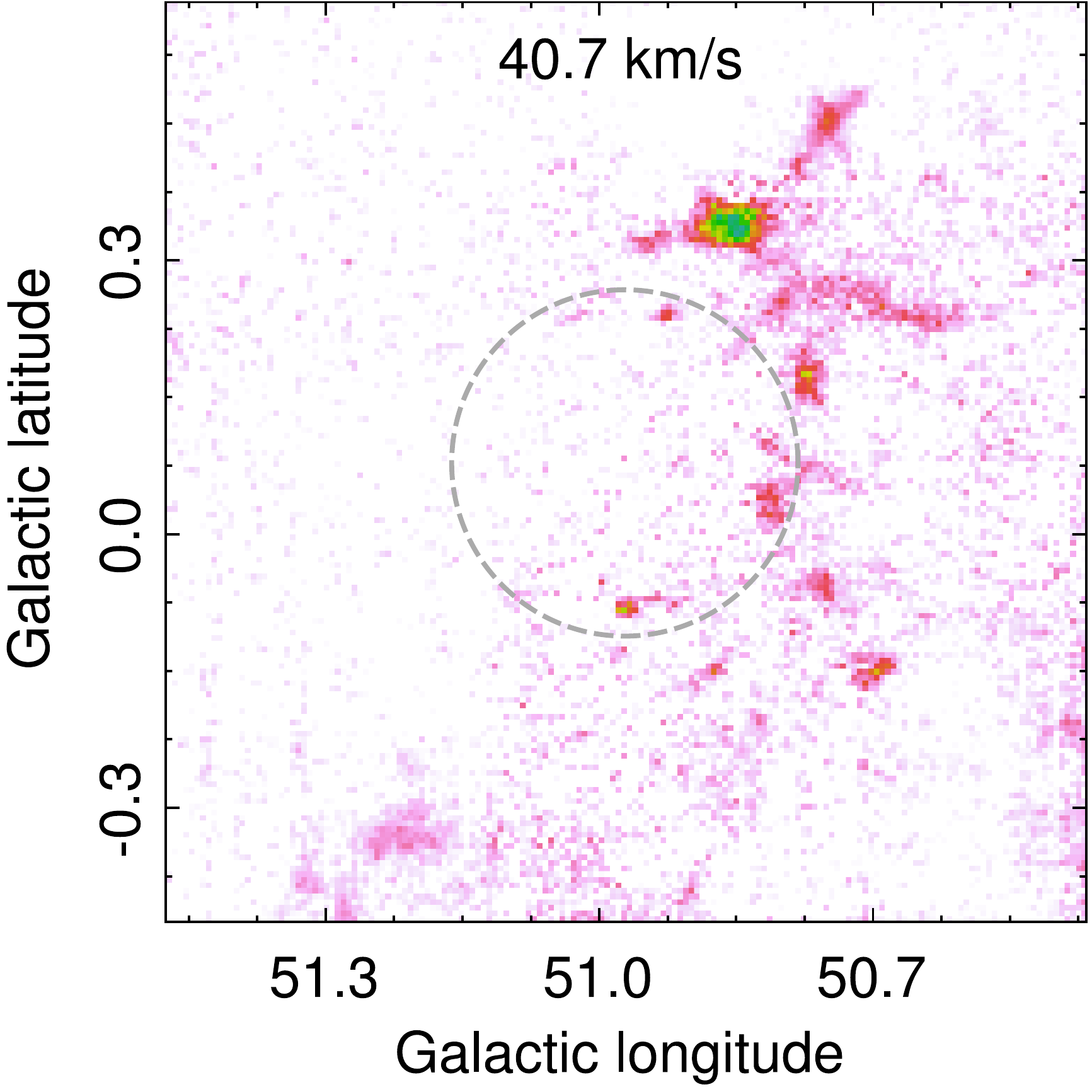}\hfill
  \includegraphics[width=.24\hsize]{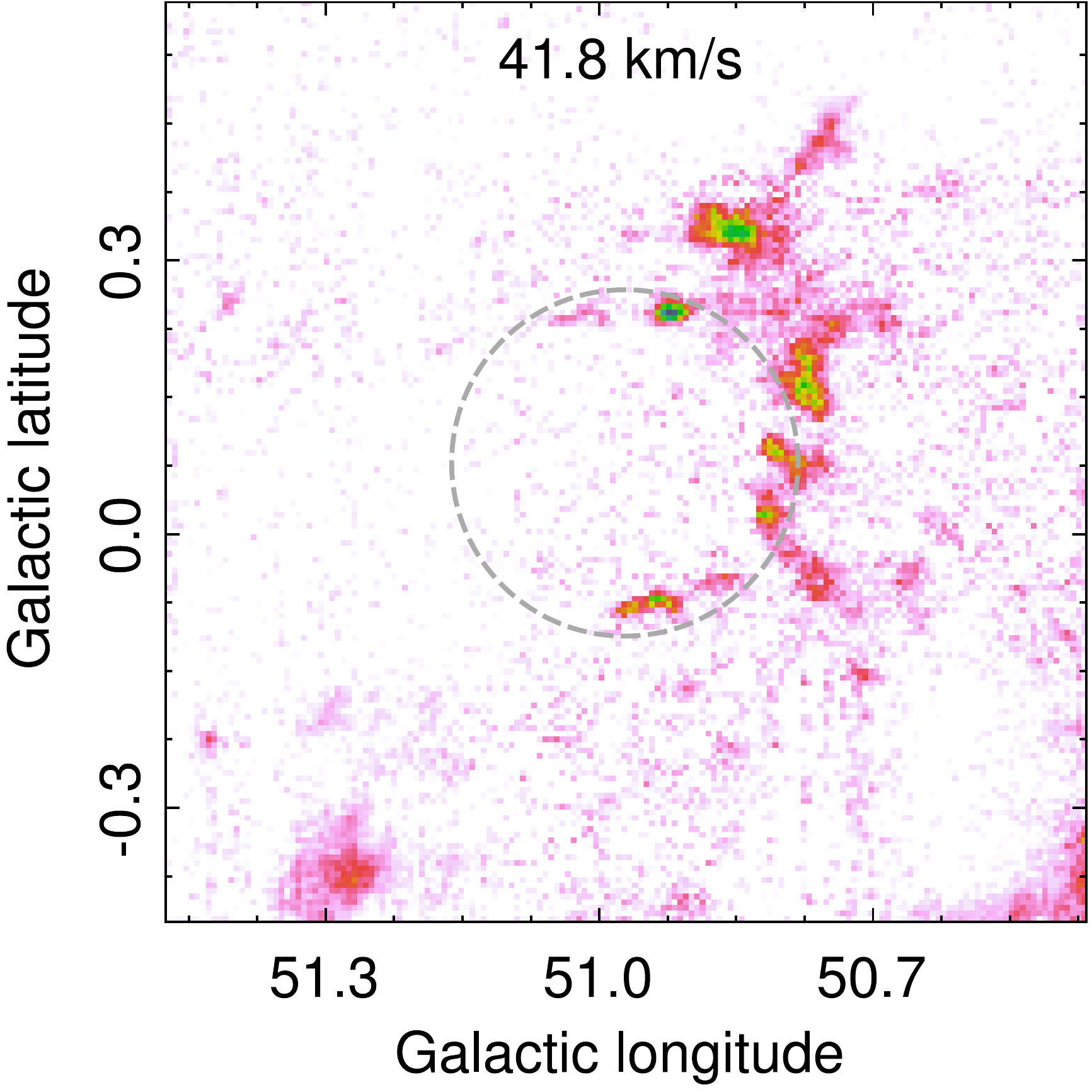}
  
  \medskip\noindent
  
  \includegraphics[width=.24\hsize]{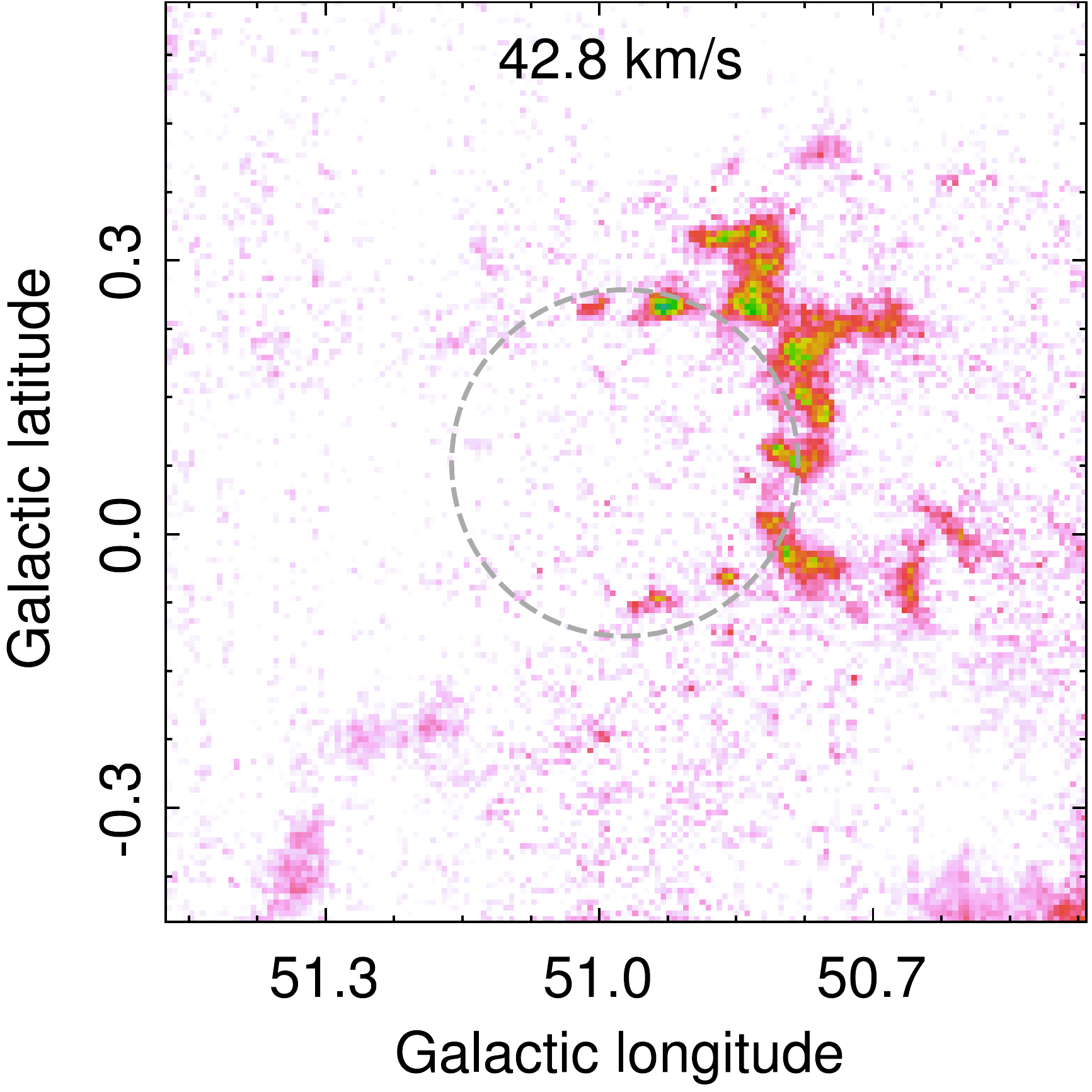}\hfill
  \includegraphics[width=.24\hsize]{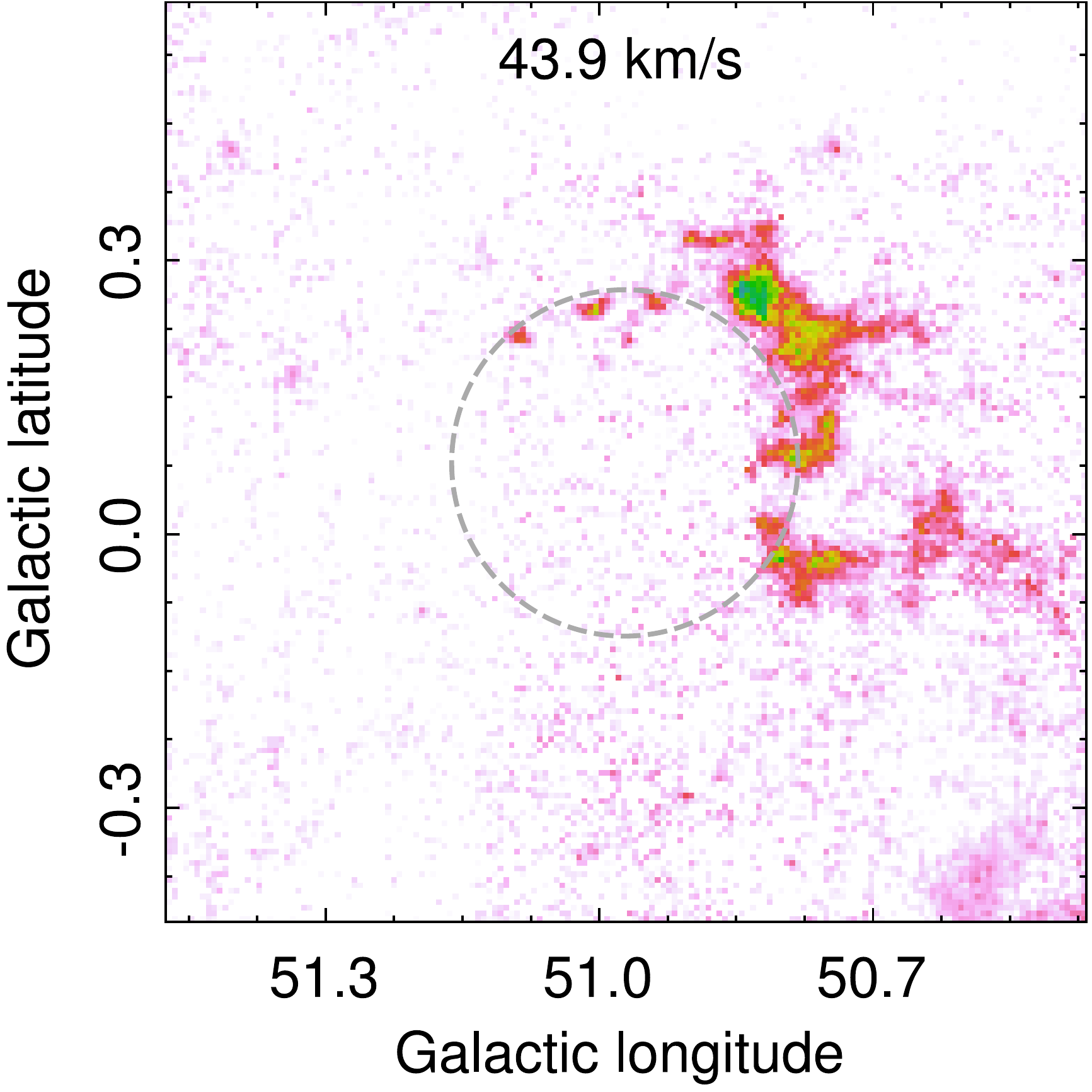}\hfill
  \includegraphics[width=.24\hsize]{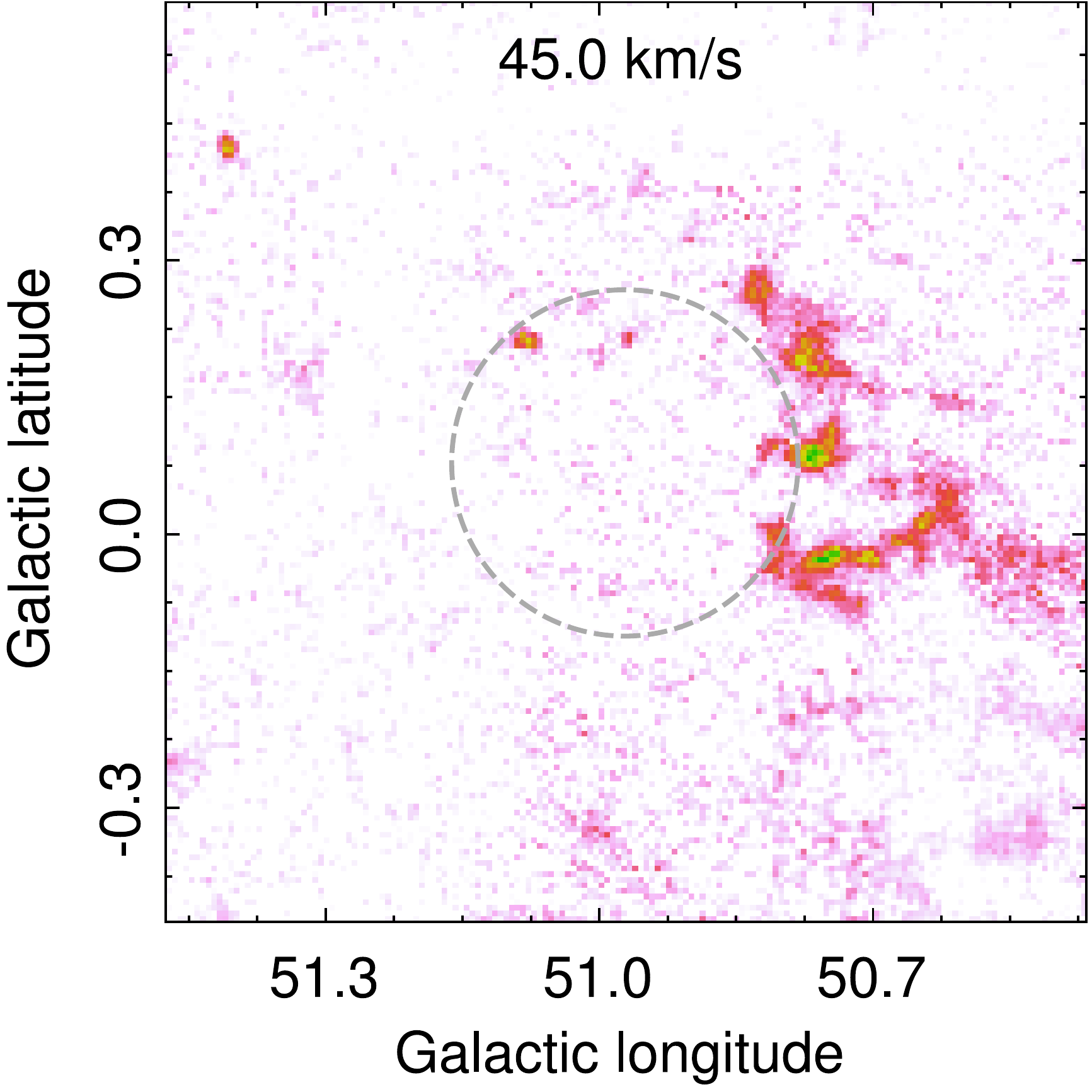}\hfill
  \includegraphics[width=.24\hsize]{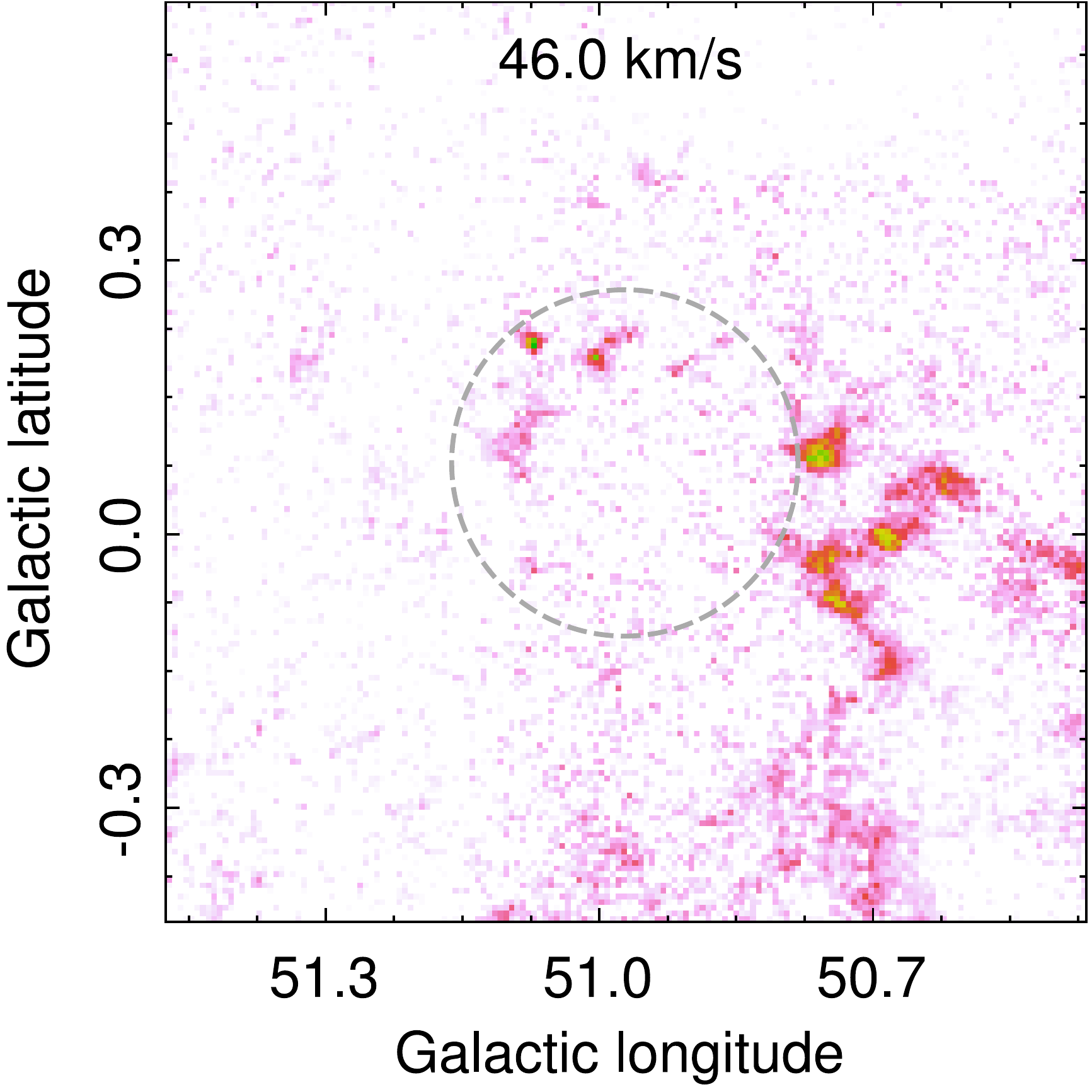}
  
  \medskip\noindent
  
  \includegraphics[width=\hsize]{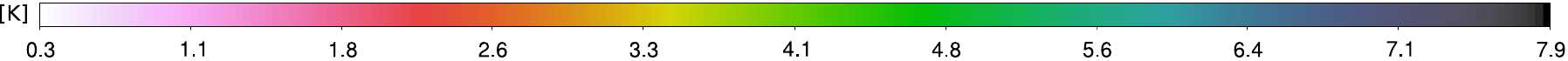}
  \caption{Brightness temperature maps of the \element[][13]{CO} ($J=1\mathrm{-}0$) emission in several velocity
    channels.  The units is kelvin.  An outline of bubble N107 as given in the catalogue of
    \citet{2006apj...649..759c} is marked with the dashed grey circle.}
  \label{fig_grs}
\end{figure*}

Around the radial velocity of $43\ \mathrm{km/s}$, CO emission is present along the bubble borders.
Maps of the CO emission (fig.~\ref{fig_grs}) show that CO gas forms clumps,
which are distributed in a ring-like structure along the edges of the bubble, protruding outside.
Almost no emission is coming from the interior.
Around the radial velocity of $48\ \mathrm{km/s}$, the CO emission merges with another complex of molecular clouds,
while in the channels below $38\ \mathrm{km/s}$, CO emission vanishes.

To measure the associated molecular mass, we use two 3D apertures (see fig.~\ref{fig_apertures}, right panel).
Aperture 1 is same as that for measuring the \ion{H}{i} mass.
Aperture 2 is smaller, with radius $r_\mathrm{ap2} = 1.36 \cdot r_\mathrm{N107} = 15\farcm5$,
radial velocity channels from $38.8$ to $46.9\ \mathrm{km/s}$ inclusive
and covers only the immediate vicinity of the bubble,
i.e. only the mass accumulated in the expanding shell.
Note that for comparing the observation and simulations,
we use the molecular mass found within aperture 2.
We treat the radiative transfer in the line of \element[][13]{CO} ($J=1\mathrm{-}0$),
described in \citet{1996tra..book.....r}.
First, we derive the optical depth:

\begin{equation}
  \label{eq_tau_13co}
  \tau = -\ln \left[ 1 - \frac{T_\mathrm{b}}{T_0} \left\{ \left[ \mathrm{e}^{T_0/T_{\mathrm{ex}}} - 1 \right]^{-1} - \left[ \mathrm{e}^{T_0/T_{\mathrm{CMB}}} - 1 \right]^{-1} \right\}^{-1} \right],
\end{equation}

\noindent
where $T_\mathrm{ex}$ is the excitation temperature, which we assume to be $20\ \mathrm{K}$.
$T_\mathrm{b}$ is the brightness temperature of the CO line;
$T_\mathrm{CMB} = 2.7\ \mathrm{K}$ is the temperature of the cosmic microwave background and

\begin{equation}
  \label{eq_tzero_13co}
  T_0 = \frac{hf}{k} \approx 5.3\ \mathrm{K},
\end{equation}

\noindent
where $h$ is the Planck constant; $k$ is the Boltzmann constant and
$f = 110\ \mathrm{GHz}$ is the \element[][13]{CO} ($J=1\mathrm{-}0$) line frequency.
Assuming local thermodynamical equilibrium, characterised by a single excitation temperature $T_\mathrm{ex}$,
we can compute the \element[][13]{CO} column density from the formula \citep{1996tra..book.....r}:

\begin{equation}
  \label{eq_n_13co}
  \frac{N_{\element[][13]{CO}}}{\mathrm{cm}^{-2}} = 2.6 \times 10^{14} \frac{T_\mathrm{ex}}{1-\mathrm{e}^{-T_0/T_\mathrm{ex}}} \int \tau(v)\,\mathrm{d}v,
\end{equation}

\noindent
where, in our case of the GRS data cube, $\mathrm{d}v = 0.21\ \mathrm{km/s}$
and the integral becomes a sum over a desired range of pixels.
From the \element[][13]{CO} column density, the H$_2$ column density is derived assuming
the abundance \citep{1998gaas.book.....b}
$N_{\mathrm{H}_2} = 6.5 \times 10^5 N_{\element[][13]{CO}}$.
Finally, the total mass of the molecular matter $m_\mathrm{mol}$ is computed using the standard solar
hydrogen abundance $X$:

\begin{equation}
  \label{eq_mmol}
  m_\mathrm{mol} = \frac{m_{\mathrm{H}_2}}{X}, \qquad \mathrm{where} \quad X = 0.735,
\end{equation}

\noindent
and $m_{\mathrm{H}_2}$ is the mass in the form of the molecular hydrogen.
Note that we use term \emph{molecular matter} or \emph{molecular mass} to refer to the matter found
in molecular clouds, i.e. a mixture of H$_2$, He, CO and trace amounts of other molecules,
elements and dust.
The distance to N107 ($3.6\ \mathrm{kpc}$) is discussed above in section~\ref{sec_atomic_component}. 

The total molecular mass within aperture 1 is $1.3 \times 10^{5}\ M_\odot$,
from which $9.8 \times 10^{4}\ M_\odot$ in in the form of H$_2$, rest in helium and heavier elements.
The total molecular mass within aperture 2 (in the expanding shell only) is $4.0 \times 10^{4}\ M_\odot$,
from which $2.9 \times 10^{4}\ M_\odot$ is in the form of H$_2$.
The major source of uncertainty is the \element[][13]{CO} abundance.
We used $N_{\mathrm{H}_2} = 6.5 \times 10^5 N_{\element[][13]{CO}}$ \citep{1998gaas.book.....b}.
Observations and models show, however, that the abundance varies for different clouds
and depends also on the physical conditions inside individual clumps
\citep{2007apjs..168...58s,2009a&a...503..323v,1999rpph...62..143w}.
Another source of the uncertainty we consider is the distance to N107.
Using eq.~\ref{eq_uncertainty_propagation} and
assuming the \element[][13]{CO} abundance uncertainty of $20\ \%$
and the distance uncertainty of $10\ \%$ ($0.36\ \mathrm{kpc}$),
the uncertainty of the molecular masses given above is about $30\ \%$.

\subsection{Radio Continuum and Spectral Index\label{sec_radio_continuum}}
Since an \ion{H}{ii} region was identified in the direction of the bubble
\citep[][region 576]{2003a&a...397..213p} and also the $24\ \mathrm{\upmu m}$ emission is
present inside the bubble, we expected to observe its radio continuum counterpart.

\begin{figure*}
  \includegraphics[width=.49\hsize]{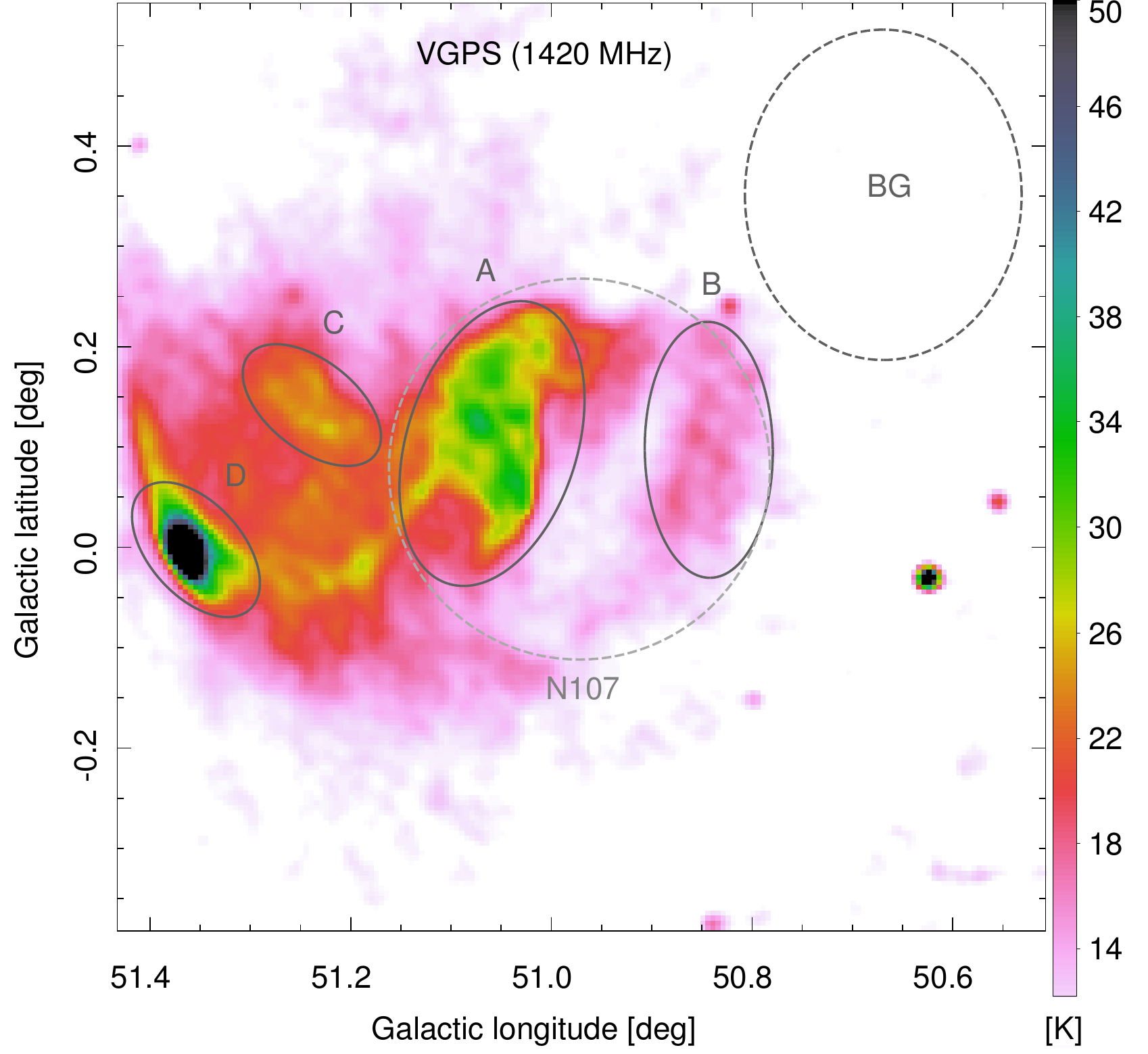}
  \includegraphics[width=.49\hsize]{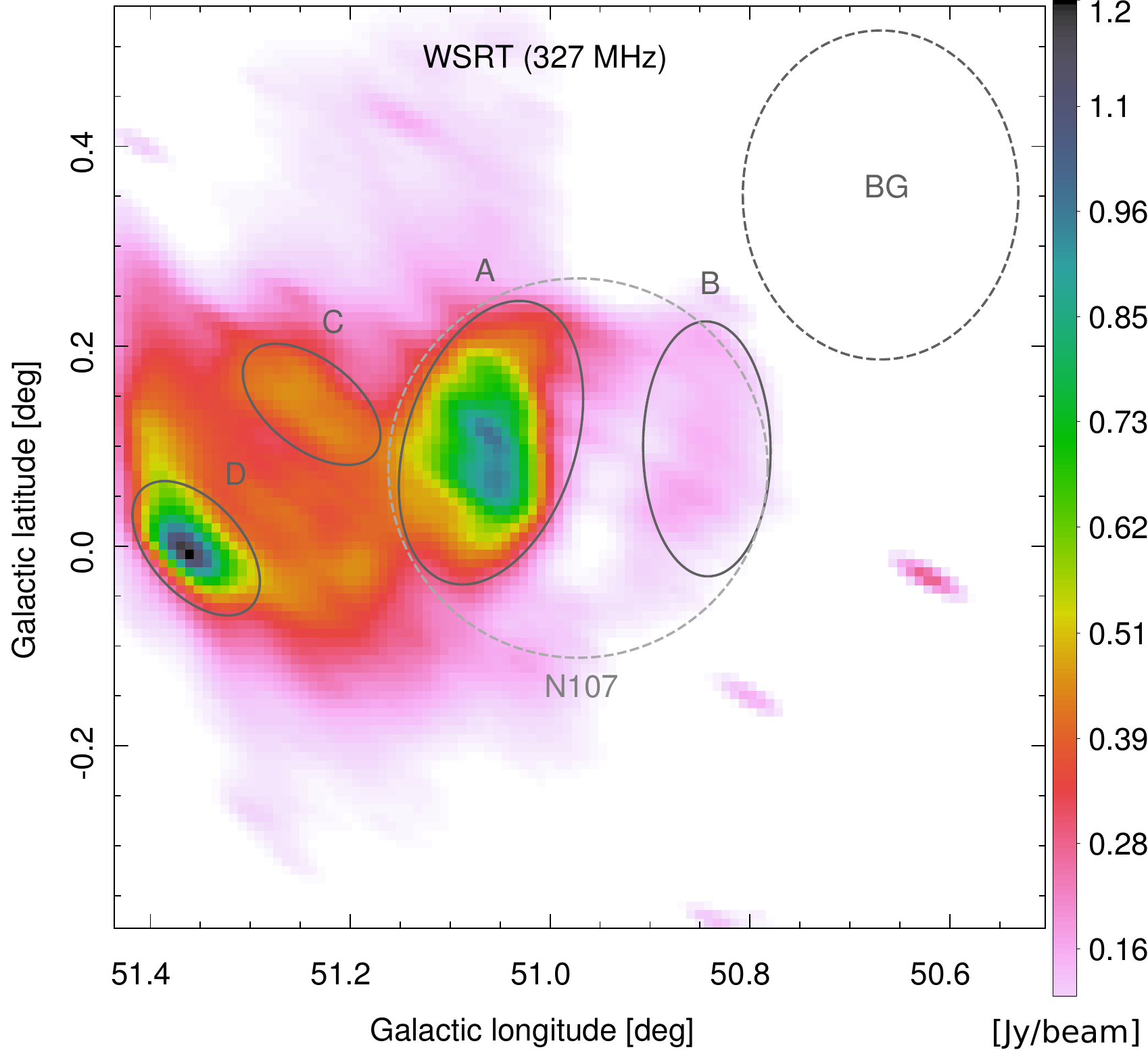}
  \caption{Radio continuum emission associated with bubble N107.
    The bubble is denoted by the grey dashed circle in the centre.
    Overlaid are four elliptical apertures (A, B, C, D) used to measure the radio fluxes.
    The grey dashed ellipse in the upper right (labelled BG) is the region used for measuring the background.
    \textbf{Left:} Emission at $1420\ \mathrm{MHz}$/$21\ \mathrm{cm}$ from the VGPS survey.
    \textbf{Right:} Emission at $327\ \mathrm{MHz}$/$92\ \mathrm{cm}$ from the WSRT survey.
    Note blurring in the direction of declination (top-left--to--bottom-right), caused by
    the larger beam size (lower resolution) in the declination direction.}
  \label{fig_radio_continuum}
\end{figure*}

\begin{figure*}
    \includegraphics[width=.49\hsize]{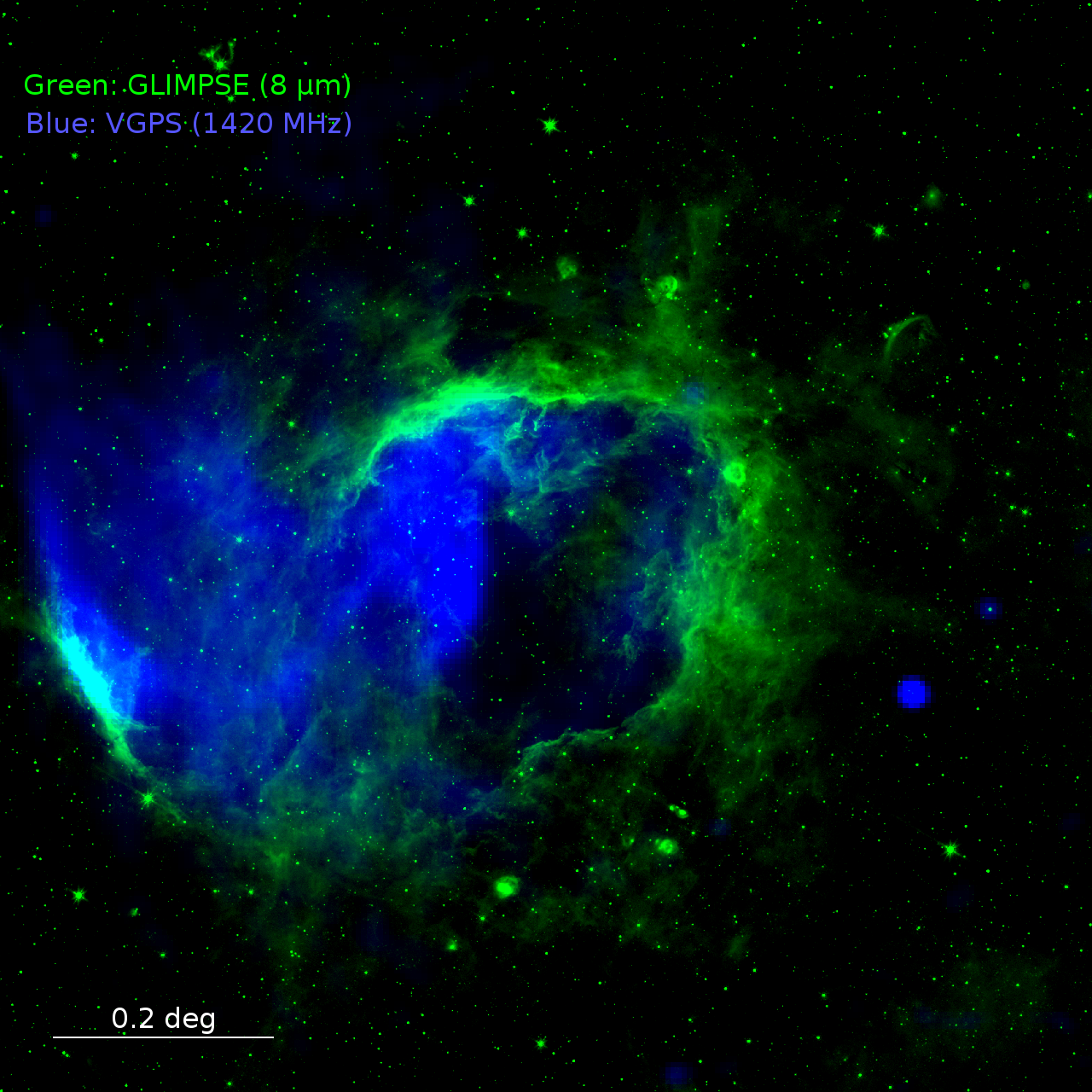}
    \includegraphics[width=.49\hsize]{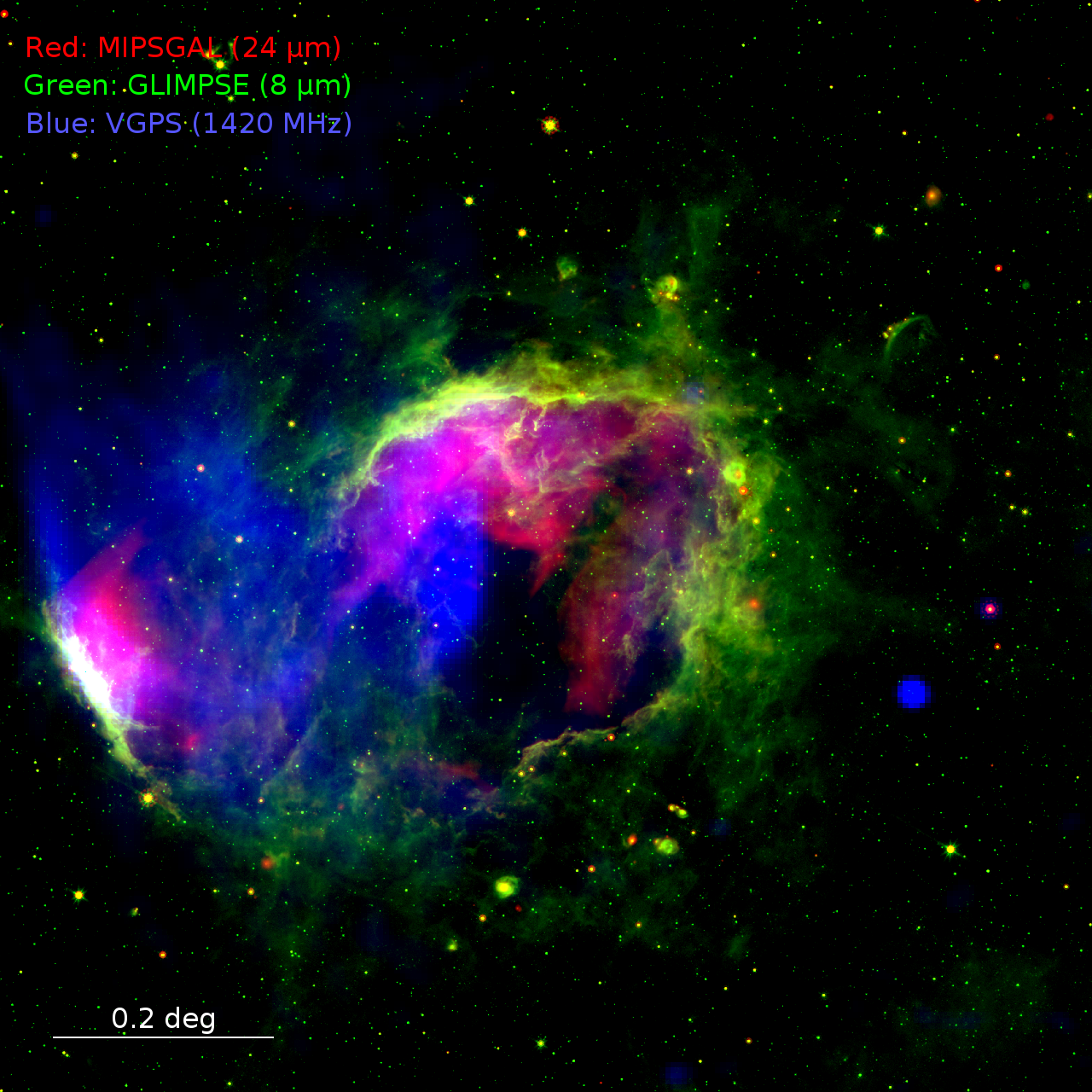}
    \caption{
        Comparison of the VGPS radio continuum at $1420\ \mathrm{MHz}$/$21\ \mathrm{cm}$ (blue) with
        the $8\ \mathrm{\upmu m}$ (green) and $24\ \mathrm{\upmu m}$ (red) emission.
        \textbf{Left:} $1420\ \mathrm{MHz}$ and $8\ \mathrm{\upmu m}$ emission.
        The radio continuum follows the bubble outline and fills part of its interior.
        \textbf{Right:} Same as left with added $24\ \mathrm{\upmu m}$ emission.
        Inside the bubble, the radio continuum is well correlated with the $24\ \mathrm{\upmu m}$ emission,
        except for the eastern part, where the $24\ \mathrm{\upmu m}$ emission is absent
        and only the radio continuum is observed.
    }
    \label{fig_vgps_multi}
\end{figure*}

Both the VGPS ($1420\ \mathrm{MHz}$) and WSRT ($327\ \mathrm{MHz}$)
surveys show a distinct diffuse radio source located apparently within the bubble,
in its eastern (left) part, slightly below the area, where the $24\ \mathrm{\upmu m}$ emission is strongest
(see fig.~\ref{fig_radio_continuum}).
Beside this source, the bubble interior is filled with much fainter radio emission.
Another diffuse radio source is present further east of the bubble ($l \approx 51\fdg35$, $b \approx 0\degr$).
This source is associated with a complex of molecular material found at radial velocities
of about $60\ \mathrm{km/s}$, so it is likely not associated with the bubble.
Fig.~\ref{fig_vgps_multi} shows a comparison of the VGPS radio continuum
with the emission at $8\ \mathrm{\upmu m}$ and $24\ \mathrm{\upmu m}$.
Inside the bubble, the radio continuum follows the bubble outline defined by the PAHs emission at $8\ \mathrm{\upmu m}$.
In the northern (top) and western (right) part,
the radio continuum is mixed with the emission of very small grains at $24\ \mathrm{\upmu m}$,
while in the eastern part (left), the $24\ \mathrm{\upmu m}$ emission vanishes and only the radio continuum is observed.

We measured the radio flux densities in four regions:
two located apparently within the bubble -- defined by apertures A and B --
and two located outside the bubble -- defined by apertures C and D
(fig.~\ref{fig_radio_continuum}).
Aperture A covers the distinct source in the eastern part of the bubble (source ``A''), while aperture B covers
the faint emission in its western (right) part (source ``B'').
Aperture C covers part of the fainter emission outside the bubble (source ``C'')
and aperture D covers the distinct radio source located eastern of the bubble (source ``D'').
We subtracted a background derived from flux
within an aperture covering a relatively homogeneous region free of visible radio sources (fig.~\ref{fig_radio_continuum}).
Then we used the two fluxes at $1420$ and $327\ \mathrm{MHz}$ to derive the spectral index $\alpha$ for both sources A and B:

\begin{equation}
  \label{eq_spindex}
  \alpha = \frac{ \ln S_1 - \ln S_2 }{ \ln f_1 - \ln f_2 },
\end{equation}

\noindent
where $S_1$, $S_2$ are the fluxes at frequencies $f_1$, $f_2$.  Note that the theoretical spectral index
for a classical \ion{H}{ii} region in the optically thin limit is $-0.1$ \citep{2010era.book......c},
although the usually observed values are around $0$ or more \citep{1996tra..book.....r,1988gera.book.....k}.
In distinction to \ion{H}{ii} regions, the mean observed spectral index of supernova remnants is $-0.5$ (average value
in catalogue of \citealt{2009ycat.7253....0g}).
The measured flux densities and the derived spectral indices are presented in table~\ref{tab_radio}.

The spectral index of source A of $-0.3$ suggests a nonthermal (synchrotron) contribution to the received
radio flux: a supernova remnant.  No known supernova remnant is associated, yet, with that radio source.
For source B, the spectral index of $\approx 0$ suggests a thermal origin of the flux:  a classical
\ion{H}{ii} region.
The spectral indices of sources C ($-0.2$) and D ($-0.1$) suggest
that part of their emission is also nonthermal.

\begin{table}
  \centering
  \caption{Radio continuum flux densities and spectral indices for apertures A, B, C and D.
    The flux densities ($S_\mathrm{A,\,B,\,C,\,D}$) are background-subtracted.}
  \label{tab_radio}
  
  \begin{tabular}{lcccc}
    \hline\hline
    Frequency &  $S_\mathrm{A}$ & $S_\mathrm{B}$ & $S_\mathrm{C}$ & $S_\mathrm{D}$ \\
    \hline
    $1420\ \mathrm{MHz}$ & $11.30\ \mathrm{Jy}$ & $2.81\ \mathrm{Jy}$ & $2.87\ \mathrm{Jy}$ & $5.12\ \mathrm{Jy}$ \\
    $\phantom{0}327\ \mathrm{MHz}$  & $17.55\ \mathrm{Jy}$ & $2.74\ \mathrm{Jy}$ & $3.91\ \mathrm{Jy}$ & $6.19\ \mathrm{Jy}$ \\
    \hline
    Spectral index & $-0.30$ & $0.02$ & $-0.21$ & $-0.13$ \\
    \hline
  \end{tabular}
  
\end{table}

%
%

\section{Analysis of Molecular Clumps\label{sec_analysis_of_molecular_clumps}}

We decomposed the molecular ring associated with N107 into individual molecular clumps
with the program DENDROFIND, described in the appendix of \citet{2012a&a...539a.116w}.
DENDROFIND processes data cubes
-- typically position-position-velocity data cubes of the brightness temperature $T_\mathrm{b}$ --
and outputs a list of
molecular clumps organised in a hierarchical structure, which can be represented by
a dendrogram \citep{2008ApJ...679.1338R}.
In our case, the data cube is a cutout of the \element[][13]{CO} GRS data
and has two spatial dimensions (galactic longitude, galactic latitude)
and one radial velocity dimension.
Each clump consists of a dense core surrounded by a less dense molecular envelope.
The core is observed as a local $T_\mathrm{b}$ peak.
DENDROFIND is controlled by four parameters:
\texttt{Nlevels} sets the number of steps by which it goes through the data cube
when searching for clump peaks;
\texttt{Npxmin} sets the minimal number of pixels which can form a clump;
\texttt{Tcutoff} sets the minimal $T_\mathrm{b}$ which is yet considered a signal,
pixels with lower $T_\mathrm{b}$ are discarded;
\texttt{dTleaf} sets the minimal height of a local peak to be considered a clump's core;
a higher \texttt{dTleaf} will merge more clumps.
Parameters \texttt{Tcutoff} and \texttt{dTleaf} are designed to be set to a value of
$\approx 3\ \times$ the noise level.
We ran DENDROFIND with the following parameters:
$\mathtt{Nlevels} = 1000$, $\mathtt{Npixmin} = 5$,
$\mathtt{Tcutoff} = \mathtt{dTleaf} = 1.0 \approx 3\sigma_\mathrm{noise}$,
where $\sigma_\mathrm{noise} \approx 0.34\ \mathrm{K}$ is the estimated standard deviation of the
gaussian noise found within our N107 cutout of the \element[][13]{CO} GRS data.
Note that this value is slightly greater then the `typical' value of the entire GRS of
$0.27\ \mathrm{K}$ given in the paper of \citet{2006apjs..163..145j}.

In order to select only the clumps associated with bubble N107, we took only those clumps
of which cores lie within the aperture 2 used for measuring the total molecular mass
(sec.~\ref{sec_molecular_component}).

Table~\ref{tab_clumps} lists the identified molecular clumps.  Note that only the
first 10 clumps with the highest core $T_\mathrm{b}$ are shown there; a complete list
is available in an online appendix (tab.~\ref{tab_clumps_full}).
The meaning of the columns is as follows:
$N$ is the number identifying a clump in our list;
$l_\mathrm{peak}$, $b_\mathrm{peak}$, $v_\mathrm{peak}$ are the galactic longitude,
galactic latitude and LSR radial velocity, respectively, of the brightest
pixel of the clump.
$T_\mathrm{peak}$ is the brightness temperature of the brightest pixel.
$R$ is the clump equivalent radius, defined as

\begin{equation}
  R = \sqrt{A_\mathrm{cl}/\pi},
\end{equation}

\noindent
where $A_\mathrm{cl}$ is the clump projected area in the sky.
$FWHM$ is the clump full width at half maximum computed as

\begin{equation}
  FWHM_x = 2\,\sqrt{2 \ln 2} \, \sqrt{ \sigma_x^2 + \left( \frac{\Delta x}{2\:\sqrt{2 \ln 2}} \right)^2 },
\end{equation}

\noindent
where $x \in \{ l,\ b,\ v \}$.
The second term under the big square root represents the pixel resolution term and
ensures that the $FWHM$ is not smaller than the pixel size.
For the GRS data cube, the pixel size $\Delta x$ is $22\farcs 1 \times 22\farcs 1 \times 0.21\ \mathrm{km/s}$,
which corresponds at N107's distance of $3600\ \mathrm{pc}$ to
$0.39\ \mathrm{pc} \times 0.39\ \mathrm{pc} \times 0.21\ \mathrm{km/s}$.
If $\Delta x$ goes to zero, the $FWHM$ approaches the gaussian $FWHM$.
$\sigma_x^2$ is the variance weighted by the brightness temperature $T_\mathrm{b}$:

\begin{equation}
  \sigma_x^2 = \overline{x^2} - \overline{x}^2
             = \sum_{\mathrm{pix}}^{} \frac{x^2 T_\mathrm{b}}{T_{\mathrm{tot}}}
               - \left( \sum_{\mathrm{pix}}^{} \frac{xT_\mathrm{b}}{T_{\mathrm{tot}}} \right)^2 .
\end{equation}

\begin{equation}
  \mathrm{where} \quad
  T_{\mathrm{tot}} = \sum_{\mathrm{pix}} T_\mathrm{b},
\end{equation}

\noindent
with the summation taken over all pixels belonging to the clump.
$M$ is the clump mass, which is the total mass of pixels belonging to the clump,
derived from equations~\ref{eq_tau_13co}--\ref{eq_mmol} (sec.~\ref{sec_molecular_component}).
The error of the clump mass, assuming gaussian noise and neglecting pixel
correlation within the beam, is

\begin{equation}
  \Delta M = \sqrt{N_\mathrm{px}} \: \Delta M_\mathrm{px},
\end{equation}

\noindent
where $N_\mathrm{px}$ is the number of pixels of the clump and $\Delta M_\mathrm{px}$
is the $\sigma_\mathrm{noise}$ equivalent mass.  In our case of the GRS \element[][13]{CO} data cubes,
$\sigma_\mathrm{noise} = 0.34\ \mathrm{K} \Rightarrow \Delta M_\mathrm{px} = 0.21\ M_\odot$.
$M_\mathrm{vir}$ is the clump virial mass, estimated as \citep{1988apj...333..821m}:

\begin{equation}
  \label{eq_mvir}
  M_\mathrm{vir} = 190 \cdot R \cdot FWHM_v^2 .
\end{equation}

\noindent
The formula above assumes a spherical clump with a density profile $\propto 1/r$.
$\Delta M_{vir}$ is the virial mass error derived from the first order Taylor expansion
of eq.~\ref{eq_mvir}, assuming the clump's area and $FWHM_v$ errors of one pixel.

\begin{figure}
   \includegraphics[width=\hsize]{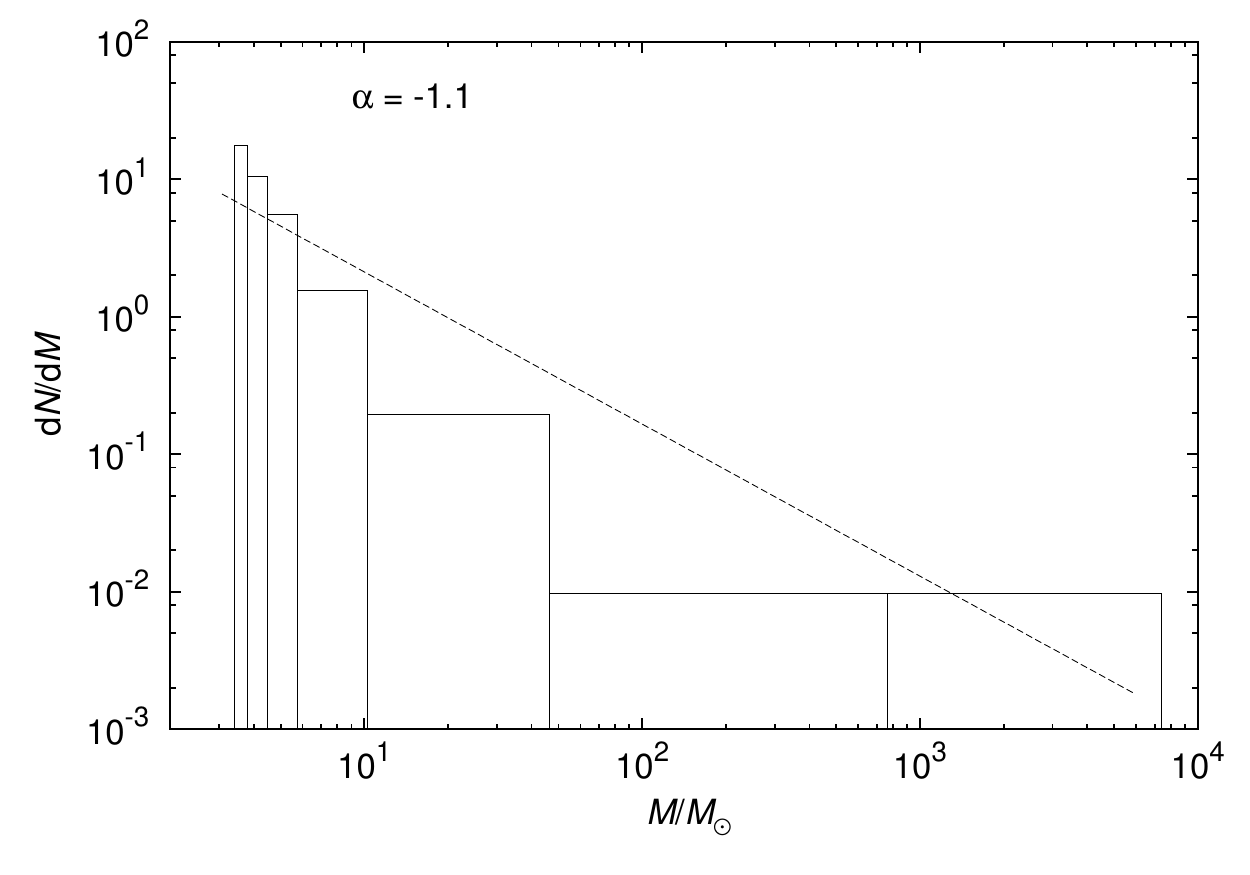}
   \caption{Histogram of the mass spectrum of molecular clumps associated with bubble N107.
     The bin widths are variable, set so that each bin contains 7 clumps.
     The dashed line represents the best fitting power-law function with the index
     $\alpha = -1.1$.
   }
   \label{fig_clumps_histogram}
\end{figure}

Fig.~\ref{fig_clumps_histogram} shows a histogram of clumps masses.  Overlaid is a line representing
a fit by a function $\mathrm{d}N/\mathrm{d}M = C (M/M_\odot)^\alpha$, where $\alpha$ is
the slope of the clump mass function and $C$ is some constant.
We used least-square fitting on a logarithmic scale, i.e.\ we fit the log of the bin heights
$\log(\mathrm{d}N/\mathrm{d}M)$ by a function $\log(C (M/M_\odot)^\alpha)$.
The best fitting slope is $\alpha = -1.1$.
The bins in the histogram have variable widths, set so that each contains 7 clumps.
This ensures that all bins have the same significance.

\begin{table*}
  \centering
  \caption{Molecular clumps associated with bubble N107.
    The first column gives a number identifying a clump in our list.
    The meaning of other columns is explained in sec.~\ref{sec_analysis_of_molecular_clumps}.
    Only the first 10 clumps are show here; a full table is available in online appendix (tab.~\ref{tab_clumps_full}).}
  \label{tab_clumps}
  
  \begin{tabular}{lrrrrrrrrrrrrr}
    \hline\hline
    $N$ &
    $\frac{l_\mathrm{peak}}{\mathrm{deg}}$ &
    $\frac{b_\mathrm{peak}}{\mathrm{deg}}$ &
    $\frac{v_\mathrm{peak}}{\mathrm{km\:s^{-1}}}$ &
    $\frac{T_{\mathrm{peak}}}{\mathrm{K}}$ &
    $\frac{R}{\mathrm{pc}}$ &
    $\frac{\mathrm{FWHM}_{l}}{\mathrm{pc}}$ &
    $\frac{\mathrm{FWHM}_{b}}{\mathrm{pc}}$ &
    $\frac{\mathrm{FWHM}_{v}}{\mathrm{km\:s^{-1}}}$ &
    $\frac{M}{M_\odot}$ &
    $\frac{\Delta M}{M_\odot}$ &
    $\frac{M_{\mathrm{vir}}}{M_\odot}$ &
    $\frac{\Delta M_{\mathrm{vir}}}{M_\odot}$ &
    $\frac{M}{M_{\mathrm{vir}}}$
    \\
    \hline
     1 & 50.919 &  0.241 &  41.89 &    7.0 &   2.2 &   2.45 &   1.62 &   1.84 & 1037.1 &    5.6 &  1409.9 &    324.8 &    0.736 \\
     2 & 50.840 &  0.266 &  44.01 &    5.8 &   4.0 &   3.48 &   3.40 &   2.83 & 4082.1 &   11.5 &  6050.2 &    909.0 &    0.675 \\
     3 & 50.753 & -0.029 &  45.08 &    4.7 &   3.2 &   3.38 &   2.52 &   4.63 & 3119.8 &   10.8 & 12847.9 &   1180.5 &    0.243 \\
     4 & 51.073 &  0.204 &  45.93 &    4.7 &   1.2 &   1.41 &   0.99 &   2.57 &  351.1 &    3.6 &  1494.4 &    247.5 &    0.235 \\
     5 & 50.766 &  0.081 &  45.50 &    4.6 &   3.3 &   3.66 &   2.21 &   3.92 & 3469.1 &   10.9 &  9622.4 &   1044.0 &    0.361 \\
     6 & 50.938 & -0.072 &  42.31 &    4.6 &   1.8 &   3.06 &   1.17 &   1.70 &  635.2 &    4.8 &  1010.1 &    252.4 &    0.629 \\
     7 & 50.778 &  0.192 &  42.74 &    4.5 &   5.0 &   5.15 &   5.07 &   2.76 & 7365.5 &   15.9 &  7259.4 &   1120.0 &    1.015 \\
     8 & 50.815 &  0.020 &  42.10 &    4.5 &   2.3 &   1.59 &   2.60 &   3.91 & 1206.7 &    6.8 &  6597.7 &    716.4 &    0.183 \\
     9 & 50.803 & -0.029 &  44.01 &    4.4 &   2.3 &   2.05 &   2.11 &   2.26 & 1059.1 &    6.1 &  2272.7 &    428.0 &    0.466 \\
    10 & 50.999 &  0.247 &  43.59 &    3.6 &   1.4 &   2.08 &   1.08 &   2.26 &  276.5 &    3.5 &  1343.3 &    252.2 &    0.206 \\
    \multicolumn{14}{l}{\dots\hfill\dots\hfill\dots} \\
    \hline
  \end{tabular}
  
\end{table*}

%
%

\section{Numerical Simulations of Bubbles and Comparison with N107\label{sec_simulations}}

We simulated the evolution of bubbles formed in the ISM around massive stars in order to estimate
parameters of N107, which cannot be derived directly from observation (esp.~its age).
We used a version of the FORTRAN code \emph{ring} \citep{1990iaus..144p.101p,1996a&a...313..478e},
which was originally written for simulations of expanding \ion{H}{i} shells in the Galactic ISM.

\subsection{Code Description\label{sec_code_description}}

The simulation starts with a small spherical shell, which is set according to
the wind-blown bubble model of \citet{1977apj...218..377w} in the pressure-driven snowplough phase.
The infinitesimally thin shell (thin-shell approximation) is divided into $nl \times np$ elements.
We solve the equations of motion including the pressure difference
between the shell interior and exterior and the mass accumulation of the ambient medium.
We do not include gravity or any other force acting on the shell.
The momentum equation for an element is

\begin{equation}
  \label{eq_ring_momentum}
  \frac{\mathrm{d}}{\mathrm{d}t}\left( m\vec{v} \right) = 
    \vec{S} (P_\mathrm{i} - P_\mathrm{e}) ,
\end{equation}

\noindent
where $m$ is the element mass, $\vec{v}$ is the element velocity, $\vec{S}$
is the element surface and $P_\mathrm{i}$ and $P_\mathrm{e}$ are the interior and exterior pressures.
The total number of elements is fixed (we used $30 \times 60$), so their surface and mass
grow with time.  The mass growth is given by the equation

\begin{equation}
  \label{eq_ring_mass_conservation}
  \frac{\mathrm{d}m}{\mathrm{d}t} = \rho (\vec{v} \cdot \vec{S}) ,
\end{equation}

\noindent
where $\rho$ is the local density of the ambient medium.

For the sake of simplicity, an energy source is represented as a continuous and constant energy input $\dot E$.
For each timestep a corresponding amount of energy is added to the bubble interior
and a new value of the interior pressure is computed as

\begin{equation}
  \label{eq_new_pi}
  P_\mathrm{i} = \frac{2}{3}\frac{E_\mathrm{i}}{V} ,
\end{equation}

\noindent
where $E_\mathrm{i}$ is the interior thermal energy and $V$ is the bubble volume.
Eq.~\ref{eq_new_pi} assumes that the interior is filled with monoatomic gas expanding adiabatically.
In the case of a shell expanding into a uniform medium,
this type of energy source gives for the shell radius, as a function of time, a power-law with an index $3/5$.
This is in between the power-law index $2/5$, given by a single (supernova) explosion
and the power-law index $4/7$, valid for an expanding \ion{H}{ii} region.

The motion of the bubble elements is integrated using the fourth-order Runge-Kutta method with adaptive timestep.

\subsection{Simulation Setup\label{sec_simulation_setup}}

\begin{figure}
  \includegraphics[width=\hsize]{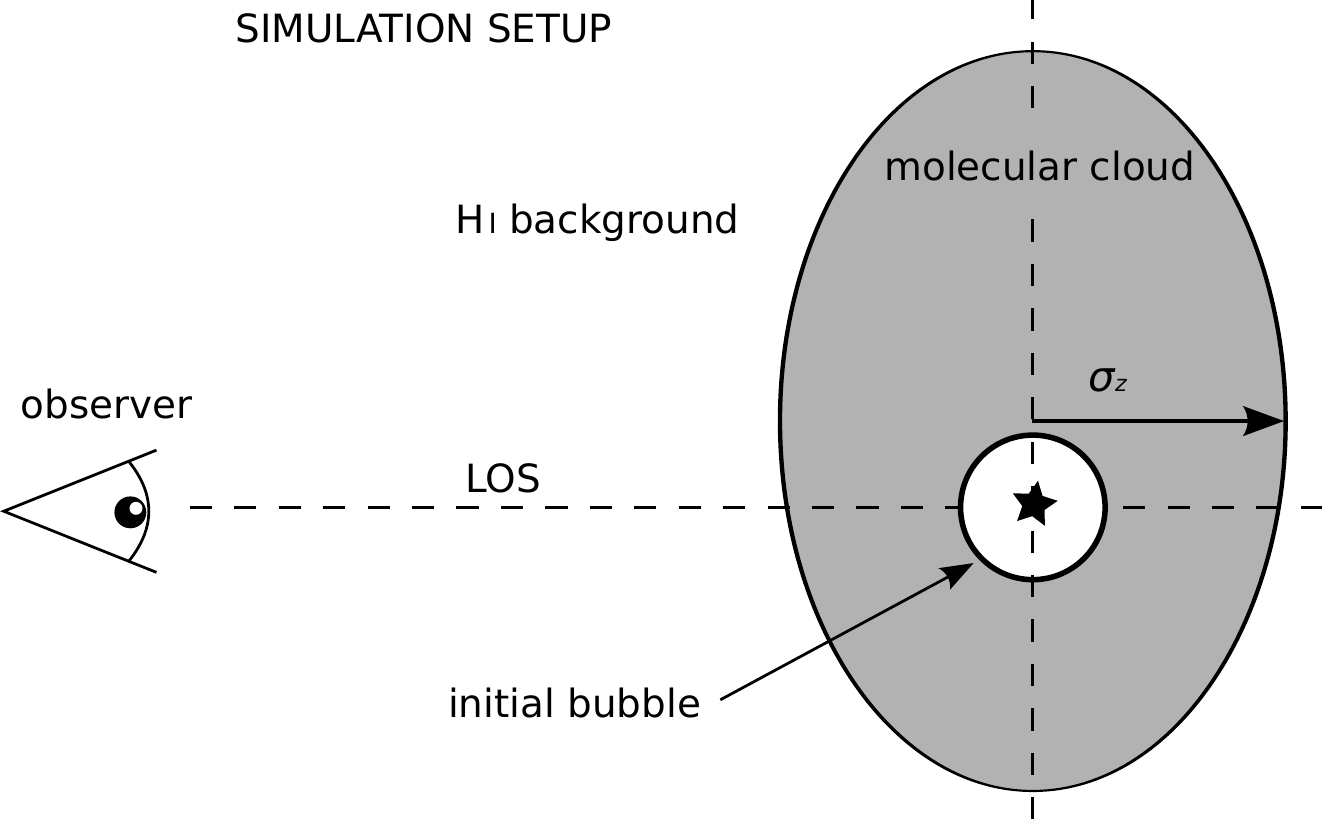}
  \caption{
    Simulation setup for a stellar-blown bubble.  The bubble is formed around a massive star(s) inside
    a molecular cloud, with a 3D gaussian density profile.
    The gaussian thickness $\sigma_x$ and $\sigma_y$, in the sky plane, is set to $12\ \mathrm{pc}$,
    while $\sigma_z$, which runs parallel to out LOS, is a parameter we vary.
    The centre of expansion is dislocated from the centre of the cloud by $\Delta r_0$, which is another
    parameter we vary.  The dislocation is towards the ($-x$, $-y$) direction, perpendicular to the LOS.
    The molecular cloud is overlaid over a background of atomic gas with the particle density of $4\ \mathrm{cm}^{-3}$.}
  \label{fig_ring_model}
\end{figure}

We assume a bubble expanding in a molecular cloud with a 3D gaussian density
profile (see fig.~\ref{fig_ring_model}):

\begin{equation}
  \label{eq_ring_density_profile}
  n = n_0 \exp \left\{ -\frac{1}{2} \left[
        \left(\frac{x-x_0}{\sigma_x}\right)^2 +
        \left(\frac{y-y_0}{\sigma_y}\right)^2 +
        \left(\frac{z-z_0}{\sigma_z}\right)^2
      \right] \right\},
\end{equation}

\noindent
where $n$ is the particle density, $n_0$ is the particle density in the centre of the cloud;
$\sigma_x$, $\sigma_y$, $\sigma_z$ are the gaussian thicknesses (standard deviation) in the $x$, $y$, $z$ direction;
$x_0$, $y_0$, $z_0$ specify the location of the cloud centre.

We also assume the centre of expansion is shifted/dislocated from the centre of the cloud by $\Delta r_0$
towards direction ($-x$, $-y$).  In the code, the expansion centre (bubble centre) is located at $x=y=z=0$,
so we shift the cloud centre by $\Delta r_0$ in the $xy$ plane:

\begin{equation}
  \label{eq_ring_shift}
  x_0 = y_0 = \Delta r_0 / \sqrt{2},  \qquad  z_0 = 0
\end{equation}

We assume a hydrogen abundance of $X = 0.735$, helium abundance of $Y = 0.245$,
metallicity of $Z = 0.02$, and a mean relative particle weight of metals of $16.65$.
We also assume that all hydrogen in the cloud is in the form of H$_2$.
This yields a mean relative particle weight of the molecular matter of $\mu = 2.34$.
The molecular material of the cloud is isothermal with the temperature set at $20\ \mathrm{K}$.
We also consider the atomic gas component, which is represented by a density background
with the constant value of $4$ hydrogen atoms per cm$^3$
and a temperature of $6000\ \mathrm{K}$.
The total initial energy of the bubble is set to $1\times 10^{49}\ \mathrm{erg}$
and each time step $\delta t$ the energy is increased by $\dot E \, \delta t$.

\subsection{Comparison with Observation\label{sec_comparison_with_observation}}

We ran a series of simulations in order to find the parameters which produce a bubble with similar mass distribution
and expansion velocity as N107.
The varying input parameters for \emph{ring} were the energy input ($\dot{E}$), the shift/dislocation of the expansion centre
from the cloud centre ($\Delta r_0$), the gaussian thickness of the cloud in the direction of our line of sight ($\sigma_z$)
and the total mass of the molecular cloud ($M_0$).
The gaussian thickness in the sky plane $\sigma_{xy} = \sigma_x = \sigma_y$ was kept constant at $12\ \mathrm{pc}$,
which is comparable to the radius of N107.
For each set of these parameters, the code evolves a bubble in time and outputs a chain of its snapshots in given time intervals.

Because the bubble is expanding in an inhomogeneous, non-spherical density field, the elements facing a lower density expand faster.
This causes the bubble's geometrical centre to move progressively towards the direction of a lower density.
To compensate for this drift, we centred each snapshot at the bubble's geometrical centre,
which is computed as the mean $(x,y,z)$ coordinates of the bubble elements weighted by their surfaces.
We projected these centred snapshots onto FITS data cubes with two spatial (XY) and one radial velocity (RV) dimensions
with the same pixel size as in the GRS data cube.

The radiation from OB stars illuminating molecular clouds creates a layer of dissociated matter.
This layer then serves as a shield against the radiation and prevents it from penetrating further inside.
The dissociated matter will not contribute to the \element[][13]{CO} line brightness temperature.
To account for that, before computing the \element[][13]{CO} line brightness temperature,
we decreased the surface density of each bubble element by $4.35\ M_\odot\, \mathrm{pc}^{-2}$.
This value is based on the suggestion for
shielding by \citet{2009apj...693..216k}.
We assume that the \element[][13]{CO} abundances are the same as we used for measuring the observed
molecular mass in sec.~\ref{sec_molecular_component}.
Then we derived for each pixel the corresponding brightness temperature
of the \element[][13]{CO} ($J=1\mathrm{-}0$) line.
Finally, we applied convolution to simulate the FCRAO telescope resolution used in the GRS.
The resulting FITS thus represents an image, which would be seen by FCRAO if it observed the simulated
bubble at a given evolution time.


We define a goodness-of-fit parameter $\delta$, to compare the synthetic data cubes to the observations:

\begin{equation}
  \label{eq_delta}
  \delta = \frac{1}{3}\left[ \delta V_\mathrm{exp}^2 + \frac{1}{4}\sum_{i=1}^{4}\left( \delta m_i^2 + \delta r_i^2 \right) \right],
\end{equation}

\begin{figure*}
    \includegraphics[width=.32\hsize]{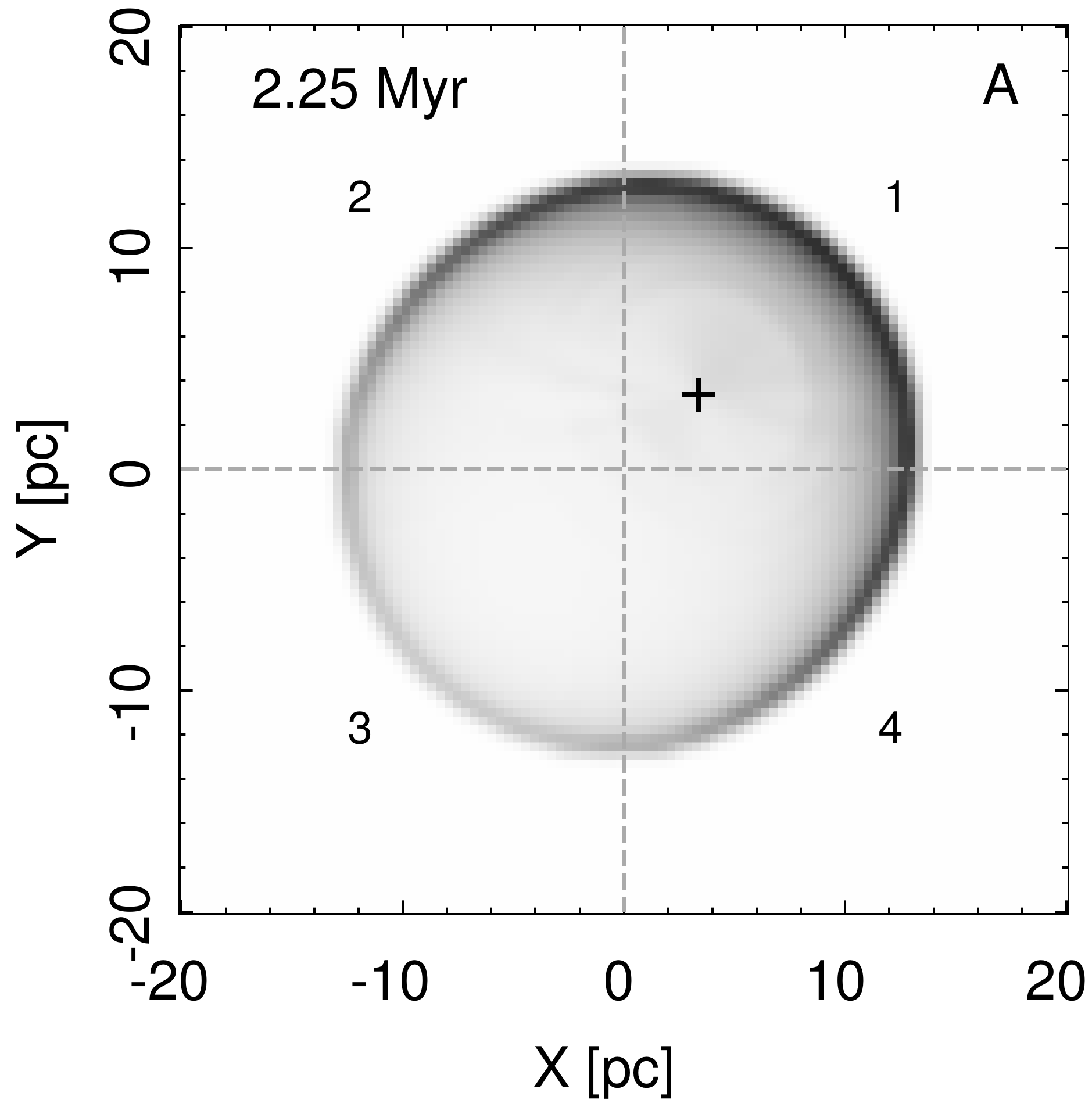}\hfill
    \includegraphics[width=.32\hsize]{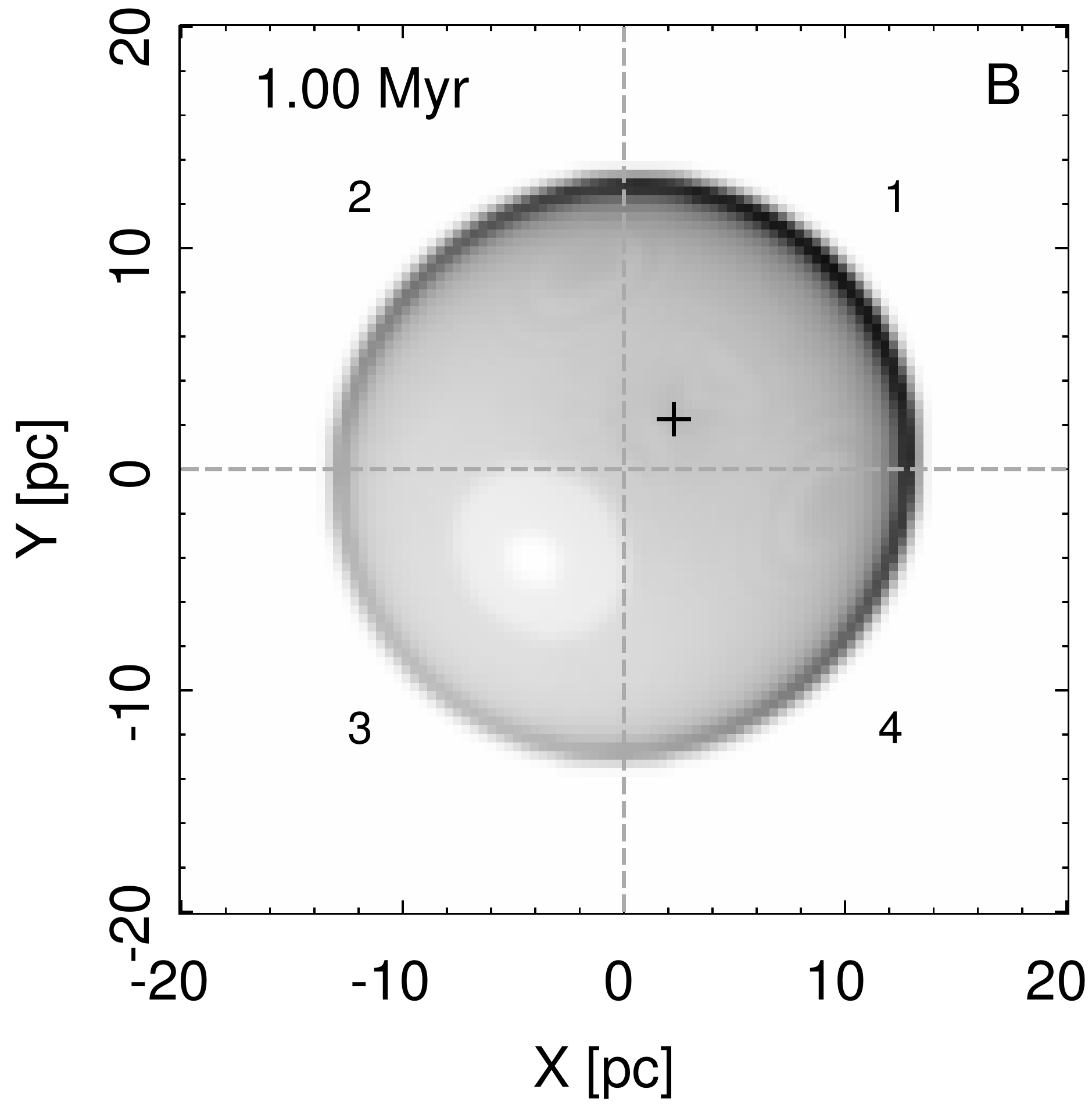}\hfill
    \includegraphics[width=.32\hsize]{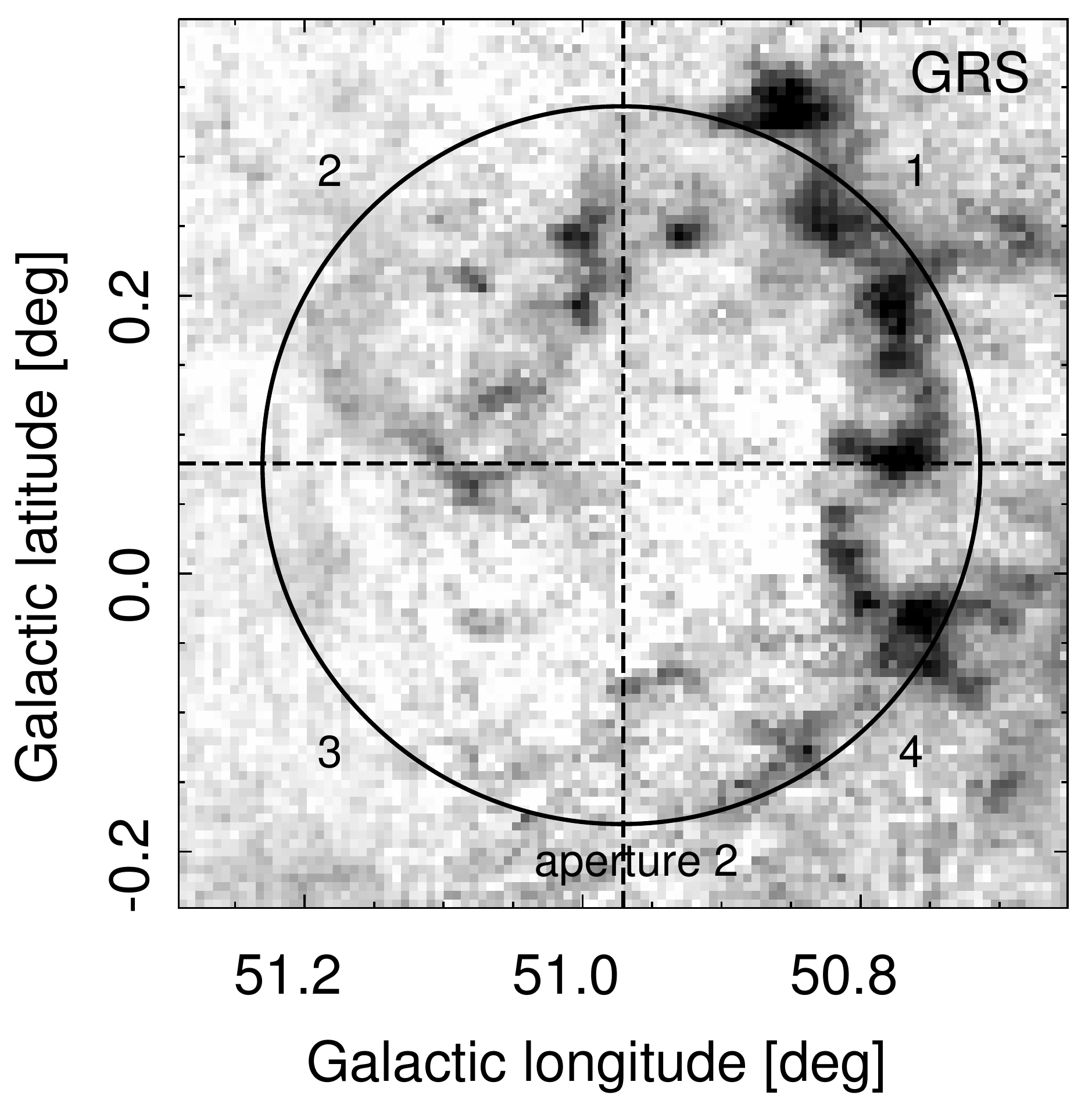}
    
    \medskip\noindent
    
    \includegraphics[width=\hsize]{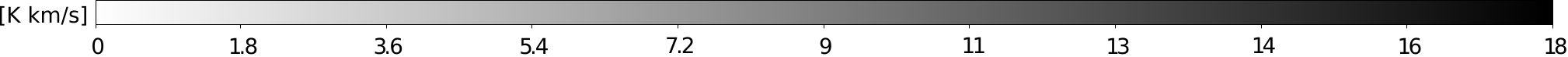}\hfill
    
    \caption{Snapshots from the best fitting simulations from group A (left) and B (centre) and GRS observation (right).
             Shown is the brightness temperature of the \element[][13]CO line integrated over the radial velocities
             of $\pm 8\ \mathrm{km/s}$ relative to the radial velocity of the bubble centre.
             The images also show four sectors used for comparing the angular mass distribution
             (see sec.~\ref{sec_comparison_with_observation} and eq.~\ref{eq_circular_sectors_sum}).
             The plus sign (+) in pictures A and B marks the original expansion centre.
             Aperture 2, shown in the GRS picture, marks which region from the observed data we used for comparison with simulations.
             }
    \label{fig_snaps}
\end{figure*}

\noindent Lower values of $\delta$ mean a better fit to (better agreement with) the observed properties of N107.
The meaning of the terms is as follows:

\begin{equation}
  \delta V_\mathrm{exp}^2 = \left( \frac{V_\mathrm{exp,\,s} - V_\mathrm{exp,\,o}}{V_\mathrm{exp,\,o}} \right)^2
\end{equation}

\noindent
represents the square of the difference of the bubble expansion velocity between the observed value
of N107 ($V_\mathrm{exp,\,o}$) and simulated value ($V_\mathrm{exp,\,s}$).
We adopt the observed bubble expansion velocity of $V_\mathrm{exp,\,o} = \ 8\ \mathrm{km/s}$, which is derived from
the \ion{H}{i} observations and represents half of the relative velocity of the shell's front and back wall.

\begin{equation}
  \label{eq_circular_sectors_sum}
  \frac{1}{4}\sum_{i=1}^{4}\left( \delta m_i^2 + \delta r_i^2 \right) =
  \frac{1}{4}\sum_{i=1}^{4}\left( \left[ \frac{m_{i,\,\mathrm{s}} - m_{i,\,\mathrm{o}}}{m_{i,\,\mathrm{o}}} \right]^2
                                 +\left[ \frac{r_{i,\,\mathrm{s}} - r_{i,\,\mathrm{o}}}{r_{i,\,\mathrm{o}}} \right]^2 \right)
\end{equation}

\noindent
is a sum taken over four circular sectors of the aperture 2 used for measuring the bubble molecular mass
(see sec.~\ref{sec_molecular_component}).
This term compares the angular mass distribution between N107 and the simulated bubbles.
A visualisation of the sectors is show in fig.~\ref{fig_snaps}.
$m_{i,\,\mathrm{s}}$ and $m_{i,\,\mathrm{o}}$ are the simulated (``s'') and observed (``o'') molecular
masses in circular sector $i$.  $r_{i,\,\mathrm{s}}$ and $r_{i,\,\mathrm{o}}$ give the simulated and observed
mean radii in sector $i$, weighted by the brightness temperature:

\begin{equation}
  r_i = \langle r_i \rangle_{T_\mathrm{b}} = \frac{\sum\limits_{(px\ \mathrm{in}\ i)} r \cdot T_\mathrm{b} }{\sum\limits_{(px\ \mathrm{in}\ i)} T_\mathrm{b} },
\end{equation}

\noindent where the sums are taken over the pixels in sector $i$.

\begin{table}
  \centering
  \caption{Ranges of the input parameters $\dot E$, $r_0$, $\sigma_z$, $M_0$ and the constraint of the evolution time ($t$)
    we used for the numerical simulations of expanding bubbles.}
  \label{tab_runs_params}
  
  \begin{tabular}{ll}
    \hline\hline
      Range of values & Steps \\
    \hline
    \\
    \hline
      \multicolumn{2}{l}{First series (wider parameter ranges)} \\
    \hline
      $\dot E = (0.1 \dots 100) \times 10^{50}\ \mathrm{erg/Myr}$  & 10 log steps \\
      $r_0 = (0.1 \dots 30)\ \mathrm{pc}$                          & 10 log steps \\
      $\sigma_z = (1 \dots 30)\ \mathrm{pc}$                       & 10 log steps \\
      $M_0 = (1 \times 10^5 \dots 1 \times 10^7)\ M_\odot$         & 10 log steps \\
      $t = (0 \dots 100)\ \mathrm{Myr}$                            & $0.5\ \mathrm{Myr}$ steps \\
    \\
    \hline
      \multicolumn{2}{l}{Second series (finer)} \\
    \hline
      $\dot E = (0.1 \dots 10) \times 10^{50}\ \mathrm{erg/Myr}$  & 8 log steps \\
      $r_0 = (10 \dots 28)\ \mathrm{pc}$                          & 10 linear steps \\
      $\sigma_z = (2 \dots 28)\ \mathrm{pc}$                      & 14 linear steps \\
      $M_0 = (3 \times 10^5 \dots 3 \times 10^6)\ M_\odot$        & 8 log steps \\
      $t = (0 \dots 8)\ \mathrm{Myr}$                             & $0.25\ \mathrm{Myr}$ steps \\
    \hline
  \end{tabular}

\end{table}

First, we ran a series of simulations with a wider span of the input parameters $\dot E$, $r_0$, $\sigma_z$, $M_0$
with steps on a logarithmic scale and the evolution time up to $t = 100\ \mathrm{Myr}$ with snapshots taken
every $0.5\ \mathrm{Myr}$.  Then we located a subset of the input parameters with best fits to the observation (lowest $\delta$)
and ran a second series of simulations with finer parameter resolutions.  In the second series, we
used linear steps for parameters $r_0$ and $\sigma_z$ and logarithmic steps for $\dot E$ and $M_0$.
The snapshots were taken every $0.25\ \mathrm{Myr}$ up to $t = 8\ \mathrm{Myr}$.
Note that the internal time step of \emph{ring} is controlled dynamically by its Runge-Kutta routines;
we only set the snapshot printout interval $\Delta t$.
Parameters for both runs are listed in table~\ref{tab_runs_params}.
The ten best fitting runs are shown in table~\ref{tab_bestfits}.

\begin{table*}
  \centering
  \caption{First ten sets of parameters best fitting the observation of N107.
           $m_{1,2,3,4}$ are the masses of four sectors of the bubble.
           $m_1/m_3$ is the ratio of masses from the first and third sector,
           which quantify the mass contrast between the most and least massive sector.
           $\delta$ is a goodness-of-fit parameter (lower is better).
           The simulations are divided into two groups A and B, discussed at the end of sec.~\ref{sec_comparison_with_observation}.
           Further explanation of the table columns is also given in sec.~\ref{sec_comparison_with_observation}.
           The asterisk (*) marks two simulations -- best of their group -- which are visualised in fig.~\ref{fig_snaps}.
          }
  \label{tab_bestfits}
  
  \begin{tabular}{ccccccccccc}
    \hline\hline
      $\frac{\dot E}{10^{50}\ \mathrm{erg/Myr}}$ &
      $\frac{r_0}{\mathrm{pc}}$ &
      $\frac{\sigma_z}{\mathrm{pc}}$ &
      $\frac{M_0}{10^5\ M_\odot}$ &
      $\frac{t}{\mathrm{Myr}}$ &
      $\frac{V_\mathrm{exp}}{\mathrm{km/s}}$ &
      $\frac{m_1/m_2/m_3/m_4}{10^3\ M_\odot}$ &
      $\frac{m_1}{m_3}$ &
      $\frac{\delta}{10^{-2}}$ &
      group \\
    \hline
      \\
      \hline
      \multicolumn{10}{l}{N107 observation} \\
      \hline
      & & & & & $8.00$ & $19.0\,/\,5.3\,/\,2.5\,/\,12.6$ & $7.6$ & & \\
      \\
      \hline
      \multicolumn{10}{l}{Ten best fitting simulations} \\
      \hline
      $0.2$ & $18$ & $6$ & $5.8$ & $2.25$ & $7.85$ & $14.8\,/\,6.8\,/\,2.2\,/\,6.8$ & $6.7$ & $3.42$ & A* \\ 
      $0.2$ & $16$ & $6$ & $4.2$ & $2.25$ & $7.90$ & $14.0\,/\,6.8\,/\,2.5\,/\,6.8$ & $5.6$ & $3.44$ & A \\ 
      $2.7$ & $12$ & $28$ & $15.5$ & $1.00$ & $8.40$ & $16.3\,/\,7.4\,/\,2.4\,/\,7.4$ & $6.8$ & $3.65$ & B* \\ 
      $2.7$ & $12$ & $20$ & $11.2$ & $1.00$ & $8.55$ & $16.2\,/\,7.4\,/\,2.4\,/\,7.4$ & $6.8$ & $3.73$ & B \\ 
      $2.7$ & $12$ & $26$ & $15.5$ & $1.00$ & $8.25$ & $16.7\,/\,7.8\,/\,2.6\,/\,7.8$ & $6.4$ & $3.82$ & B \\ 
      $2.7$ & $10$ & $26$ & $11.2$ & $1.00$ & $8.20$ & $14.5\,/\,7.2\,/\,2.8\,/\,7.2$ & $5.2$ & $3.85$ & B \\ 
      $2.7$ & $10$ & $28$ & $11.2$ & $1.00$ & $8.35$ & $14.1\,/\,6.8\,/\,2.5\,/\,6.8$ & $5.6$ & $3.88$ & B \\ 
      $0.4$ & $22$ & $6$ & $15.5$ & $1.75$ & $8.50$ & $17.1\,/\,7.6\,/\,2.1\,/\,7.6$ & $8.1$ & $3.89$ & A \\ 
      $0.2$ & $20$ & $6$ & $8.0$ & $2.25$ & $8.05$ & $15.8\,/\,6.5\,/\,1.7\,/\,6.5$ & $9.3$ & $3.96$ & A \\ 
      $2.7$ & $10$ & $20$ & $8.0$ & $1.00$ & $8.50$ & $14.0\,/\,6.8\,/\,2.6\,/\,6.8$ & $5.4$ & $3.96$ & B \\ 
      
    \hline
  \end{tabular}
  
\end{table*}

To conclude, we simulated the evolution of bubbles formed around massive stars in molecular
clouds and compared their properties to N107.  We got the best agreement for two groups
of parameters:

\begin{enumerate}
\item Group A represents bubbles produced with a lower energy input ($\dot E < 0.4 \times 10^{50}\ \mathrm{erg/Myr}$),
greater centre-of-expansion dislocation ($r_0 \approx 16 \dots 22\ \mathrm{pc}$),
formed within an oblate molecular cloud with the oblateness of ($\sigma_z/\sigma_{xy} = 6/12$),
and with a greater evolution time ($t \approx 1.75 \dots 2.25\ \mathrm{Myr}$).

\item Group B represents bubbles produced with a higher energy input ($\dot E = 2.7 \times 10^{50}\ \mathrm{erg/Myr}$),
smaller centre-of-expansion dislocation ($r_0 \approx 10 \dots 12\ \mathrm{pc}$),
formed within a prolate molecular cloud ($\sigma_z/\sigma_{xy} \approx (20 \dots 28)/12$),
with a smaller evolution time ($t = 1\ \mathrm{Myr}$).
\end{enumerate}

\noindent
Fits show little sensitivity to the parameter $M_0$ -- the initial mass of the surrounding molecular cloud
($M_0 \approx 5.8 \dots 15.5 \times 10^5\ M_\odot$).
Both groups of the best-fittings-simulations suggest that bubble N107 is relatively young
($t \lessapprox 2.25\ \mathrm{Myr}$) and was formed at the very edge of its parental molecular cloud.

Fig.~\ref{fig_snaps} shows snapshots from the best fitting simulations from both groups A and B
and the GRS observation of the CO line.

%
%

\section{Discussion\label{sec_discussion}}

\subsection{Numerical Simulation Setup, Assumptions and Validity of Approximations}

An important assumption for the simulation setup is the shape and density profile
of the parental molecular cloud.  In our simulations, we chose the gaussian density profile and
tested different levels of the oblateness (ratio of $\sigma_z$ to $\sigma_{xy}$).  The oblate shape of the parental
molecular clouds which host bubbles similar to N107 is supported by a study of \citet{2010apj...709..791b},
who concluded that the Spitzer bubbles are formed in oblate, sheet-like, molecular clouds with
a thickness of a few parsecs.  Our simulations, however, favour neither oblate nor prolate clouds.

The numerical code \emph{ring} assumes the thin-shell approximation, which is applicable so long as the bubble
expansion is supersonic, i.e.\ expanding with a velocity higher than the sound speed of the ambient cold gas:
$\approx 0.3\ \mathrm{km/s}$ (for $T = 20\ \mathrm{K}$, $\gamma = 7/5$, $\mu = 2.34$).
In our runs, almost all the bubble's elements move faster through the ambient gas during the whole
bubble evolution, so the thin-shell approximation is reasonable.  Moreover, if the speed of an element
decreases below the sound speed, the code stops the mass accumulation for that element.

\subsection{Source of Radio Continuum\label{sec_source_of_radio_continuum}}

The radio continuum emission coming from the direction of the bubble interior has both thermal and
nonthermal components.  While the strong radio source A is dominated by nonthermal radiation,
fainter source B is emitting thermally (fig.~\ref{fig_radio_continuum}).
The nonthermal source A is brighter than the thermal source B by almost one order of magnitude,
so we can expect that the thermal emission is present in the whole bubble volume, only outshone
in aperture A by the nonthermal source.
Mixing of thermal and nonthermal radiation in aperture A is also suggested by the visual comparison of
the radio continuum and the $24\ \mathrm{\upmu m}$ emission (fig.~\ref{fig_vgps_multi}).
The northern part of aperture A features both the radio continuum and the $24\ \mathrm{\upmu m}$ emission,
while in the southern part only the radio continuum is observed.
Further, merging of thermal and nonthermal radiation
in aperture A explains why the measured spectral index ($-0.3$) is closer to $0$ than
the typical value for supernova remnants ($-0.5$, the average value in the SNRs catalogue of \citealp{2009ycat.7253....0g}).

The negative spectral indices of sources C ($-0.2$) and D ($-0.1$) suggest that
part of their radio continuum is also emitted by a nonthermal source.
The nonthermal contribution to the flux is stronger for source C than for source D.
This is expected, since aperture D covers a distinct $24\ \mathrm{\upmu m}$ (thermal) source,
while there is almost no $24\ \mathrm{\upmu m}$ emission in aperture C (fig.~\ref{fig_vgps_multi}).

Given that first supernovae in an OB association explode only after $3\ \mathrm{Myr}$ or later,
and considering the bubble evolution time estimated from simulations ($< 2.25\ \mathrm{Myr}$),
no supernova remnant should be present inside the bubble, yet.
In other words, our analysis suggests that
a supernova remnant, not associated with the bubble, may be present in the direction towards the bubble.

\subsection{Missing OB Association\label{sec_missing_ob_association}}

\begin{figure}
    \includegraphics[width=\hsize]{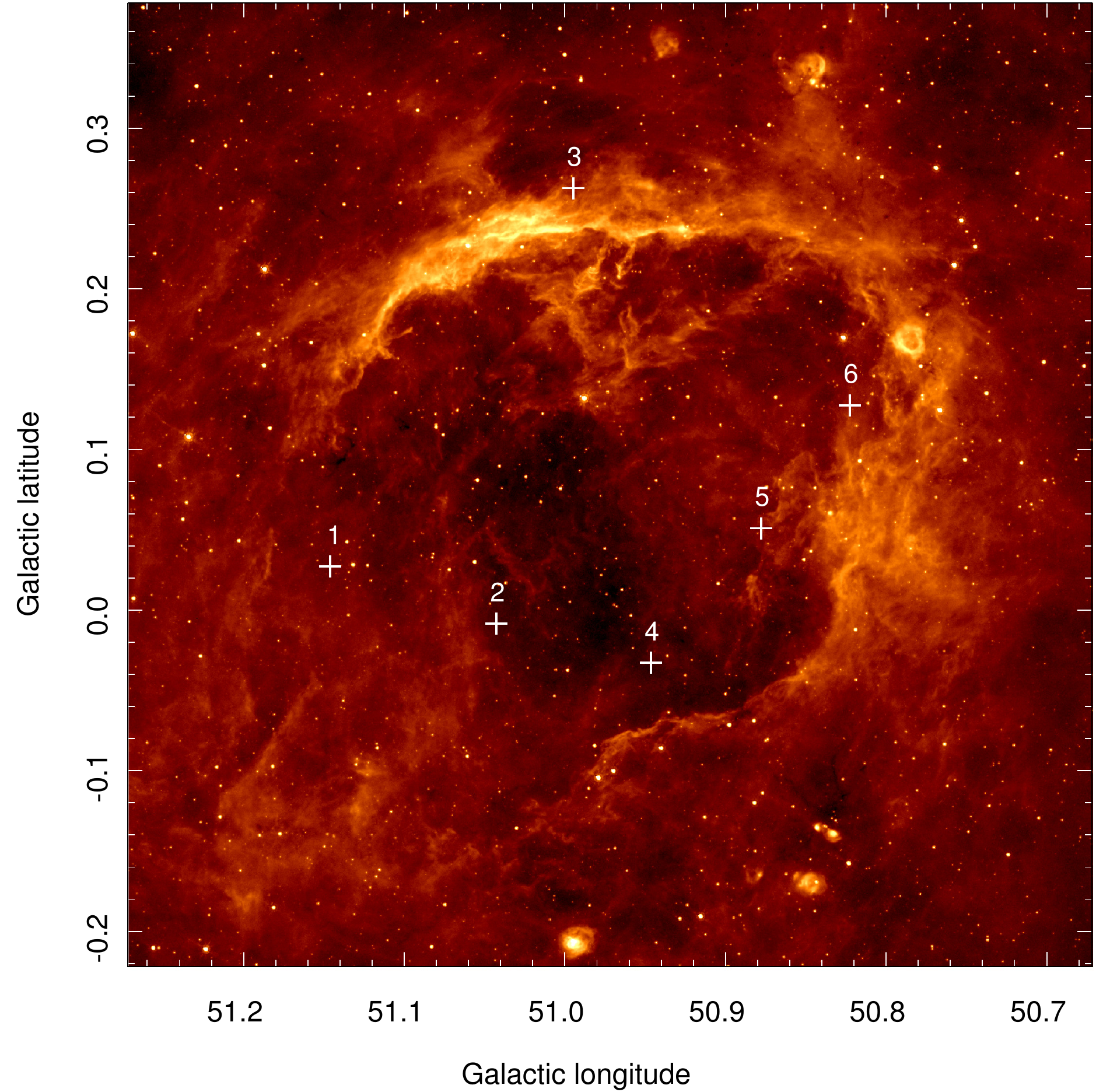}
    \caption{
        Possible stellar progenitors (white plus signs) of bubble N107 found in the UKIDSS catalogue.
        The numerical labels of the stars correspond to the labels in table~\ref{tab_ukidss_sources}.
        The background is the $8\ \mathrm{\upmu m}$ emission taken from GLIMPSE.
    }
    \label{fig_ukidss_sources}
\end{figure}

\begin{table*}
    \centering
    \caption{
        Possible stellar progenitors of bubble N107 found in the UKIDSS catalogue.
        First column gives the source label as marked in fig.~\ref{fig_ukidss_sources}.
        Second column gives the ID from the UKIDSS catalogue.
        Third and fourth columns give the galactic longitude and latitude.
        Fifth column gives the estimated mass of the source
        and sixth column gives the estimated distance.
    }
    \label{tab_ukidss_sources}
    
    \begin{tabular}{cccccc}
        \hline\hline
        label in fig.~\ref{fig_ukidss_sources} & source ID (UKIDSS) & $\frac{l}{\mathrm{deg}}$ & $\frac{b}{\mathrm{deg}}$ & $\frac{\mathrm{mass}}{M_\odot}$ & $\frac{\mathrm{distance}}{\mathrm{kpc}}$ \\
        \hline
        1 & 438326491324 & $51.1461745$ & $+0.0272455$ & $8.00$ & $3.705$ \\
        2 & 438326507390 & $51.0425744$ & $-0.0083589$ & $9.00$ & $4.044$ \\
        3 & 438326996219 & $50.9947238$ & $+0.2626358$ & $9.00$ & $4.112$ \\
        4 & 438328069418 & $50.9462654$ & $-0.0326783$ & $9.00$ & $3.193$ \\
        5 & 438328066620 & $50.8777609$ & $+0.0509439$ & $9.00$ & $3.219$ \\
        6 & 438328834208 & $50.8225691$ & $+0.1274098$ & $9.00$ & $3.232$ \\
        \hline
    \end{tabular}
    
\end{table*}

We searched the SIMBAD database and a catalogue of Galactic OB stars by \citet{2003aj....125.2531r}
for sources which could be responsible for the bubble creation.
However, no OB star, supernova remnant or pulsar is known to lie in the direction of the bubble interior or in its vicinity.

We also performed a search for massive stars possibly related to N107 using
the UKIDSS catalogue 7th data release \citep{2007MNRAS.379.1599L},
stellar evolutionary tracks by \citet{2012MNRAS.427..127B}
and the standard model of extinction \citep{1985ApJ...288..618R}.
We derived the extinction for the massive star candidates in the field of N107
and compared it to the values of extinction given for that direction by \citet{2009MNRAS.400..731S}.
The search for massive stars involves many uncertainties,
but the comparison of the derived extinctions with independent values is a good test of the credibility of our results.
We ended up with six massive star candidates, which lie at distances comparable with the distance of N107 ($3.6\ \mathrm{kpc}$).
None of these six stars have measured radial velocities.
Three of them lie at the rim of the bubble, so they can hardly be its progenitors,
but the others are possible energy sources.
The selected massive stars are shown in fig.~\ref{fig_ukidss_sources}
and listed in table~\ref{tab_ukidss_sources}.

\subsection{Molecular Clump Mass Spectrum}

The best fitting slope for the mass spectrum of molecular clumps found along the borders of N107
is $-1.1$ (fig.~\ref{fig_clumps_histogram}).
This is a significantly shallower slope than that of the classical stellar initial mass function
(Salpeter: $-2.35$) and also shallower than the clump mass function slope of $-1.6$ to $-1.8$
derived by \citet{1998a&a...329..249k}.  Note that \citet{1998a&a...329..249k} used gaussian decomposition
for clump identification, which is a fundamentally different method than we used and, moreover,
studied different types of clouds -- not shell-like as is our case of N107.

The least massive clumps we found have masses about $3.5\ M_\odot$.  This limit is given by the angular
resolution and sensitivity of the observations of the \element[][13]CO line.  With better resolution
and sensitivity, we expect that some clumps, especially the most massive ones, would be identified
as several less massive clumps.  This would alter the shape of the clump mass spectrum, and likely
redistribute the mass of the largest clumps into less massive ones, so that the slope would be steeper.

%
%

\section{Conclusion\label{sec_conclusion}}

We studied dust bubble N107, discovered by \citet{2006apj...649..759c} with infrared observations
by the Spitzer Space Telescope.
Radio observations of the \ion{H}{i} and \element[][13]{CO} ($J=1\mathrm{-}0$) lines revealed its
atomic and molecular components, which allowed us to determine the bubble radial velocity,
kinematical distance and mass of these components.
Using the code DENDROFIND \citep{2012a&a...539a.116w} we decomposed the molecular ring associated
with the bubble into 49 individual clumps and found that the clump mass spectrum slope is about $-1.1$.
Radio continuum observations at $1420$ and $327\ \mathrm{MHz}$ suggest the radio flux coming from
the bubble direction has both thermal and nonthermal components; i.e., besides a classical \ion{H}{ii}
region, a supernova remnant contributes to the emission.
We simulated the evolution of stellar-blown bubbles within molecular clouds with the numerical code \emph{ring}
\citep{1990iaus..144p.101p,1996a&a...313..478e}.
We were able to produce two groups of bubbles with a similar angular mass distribution and expansion velocity as N107:
\textbf{(A)} Bubbles formed within an oblate molecular cloud ($\sigma_z/\sigma_{xy} = 6/12$),
    with the energy input $\dot E < 0.4 \times 10^{50}\ \mathrm{erg/Myr}$ and
    the evolution time $t \approx 1.75 \dots 2.25\ \mathrm{Myr}$.
\textbf{(B)} Bubbles formed within a prolate molecular cloud ($\sigma_z/\sigma_{xy} = (20 \dots 28)/12$),
    with the energy input $\dot E \approx 2.7 \times 10^{50}\ \mathrm{erg/Myr}$ and
    the evolution time  $t = 1\ \mathrm{Myr}$.
Considering the estimated bubble evolution time ($< 2.25\ \mathrm{Myr}$),
no supernova remnant should be present inside the bubble, yet.
This may be explained by a supernova remnant present in the direction towards the bubble,
however not associated with it.
A summary of the parameters we derived for bubble N107, together with the parameters given by \citet{2006apj...649..759c},
is presented in table~\ref{tab_summary}.

\begin{table}
  \centering
  \caption{
    Summary of parameters derived for N107.
    The first section gives the values from the original catalogue of \citet{2006apj...649..759c}.
    The second section gives the values derived from observation (sec.~\ref{sec_observation}).
    The third section gives the values estimated from simulations (sec.~\ref{sec_simulations}).
    }
  \label{tab_summary}
  
  \begin{tabular}{ll}
    \hline\hline
      Property  &  Value \\
    \hline
    \\
    \hline
      \multicolumn{2}{l}{From catalogue of \citet{2006apj...649..759c}} \\
    \hline
      Galactic longitude (central)   & $50\fdg972$ \\
      Galactic latitude (central)    & $0\fdg078$ \\
      Angular radius (mean)          & $11\farcm39$ \\
      Angular thickness (mean)       & $2\farcm26$ \\
    \\
    \hline
      \multicolumn{2}{l}{Derived from observations} \\
    \hline
      Radial velocity (central)        & $43\ \mathrm{km/s}$ \\
      Expansion velocity               & $\ 8\ \mathrm{km/s}$ \\
      Distance (kinematical)           & $3.6\ \mathrm{kpc}$ \\
      Radius (mean)                    & $11.9\ \mathrm{pc}$ \\
      Mass in \ion{H}{i} (aperture 1)  & $(5.4 \pm 1.0) \times 10^3\ M_\odot$ \\
      \multicolumn{2}{l}{Mass of molecular component (aperture 1)} \\
                                       & $(1.3 \pm 0.4) \times 10^5\ M_\odot$ \\
      \multicolumn{2}{l}{Mass of molecular component (aperture 2)\tablefootmark{1}} \\
                                       & $(4.0 \pm 1.2) \times 10^4\ M_\odot$ \\
    \\
    \hline
      \multicolumn{2}{l}{Estimated from simulations} \\
    \hline
      Energy input rate              & $< 0.4 \times 10^{50}\ \mathrm{erg/Myr}$\tablefootmark{A} \\
                                     & $\approx 2.7 \times 10^{50}\ \mathrm{erg/Myr}$\tablefootmark{B} \\
      Dislocation of expansion centre & $\approx 16 \dots 22\ \mathrm{pc}$\tablefootmark{A} \\
                                      & $\approx 10 \dots 12\ \mathrm{pc}$\tablefootmark{B} \\
      Parental cloud thickness in LOS  & $\approx 6\ \mathrm{pc}$\tablefootmark{A} \\
                                           & $\approx 20 \dots 28\ \mathrm{pc}$\tablefootmark{B} \\
      Parental cloud total mass      & $\approx 5.8 \dots 15.5 \times 10^5\ M_\odot$ \\
      Evolution time (age)           & $\approx 1.75 \dots 2.25\ \mathrm{Myr}$\tablefootmark{A} \\
                                     & $\approx 1 \mathrm{Myr}$\tablefootmark{B} \\
    \hline
  \end{tabular}
  
  \tablefoot{
    \tablefoottext{1}{Aperture 2 covers only the immediate vicinity of the bubble.
    See fig.~\ref{fig_apertures} for an illustration.}
    \tablefoottext{A}{Estimated from simulations from group A.}
    \tablefoottext{B}{Estimated from simulations from group B.}
  }
  
\end{table}



\begin{acknowledgements}

We thank to the anonymous referee for multiple constructive suggestions which led to improvements of the paper.
%
This work was supported by the project RVO:~67985815.
VS acknowledges
    support from a Marie Curie fellowship as part of the European Commission FP6 Research Training Network `Constellation'
        under contract MCRTN--CT--2006-035890;
    support from Doctoral grant of the Czech Science Foundation No.~205/09/H033.
%
RW acknowledges support from the project P209/12/1795 of the Czech Science Foundation.
The research leading to these results has received funding from the European Community's Seventh Framework Programme
under grant agreement no. PIIF-GA-2008-221289.
This research has made use of SAOImage DS9, developed by Smithsonian Astrophysical Observatory \citep{2003ASPC..295..489J}.
This research has made use of the SIMBAD database, operated at CDS, Strasbourg, France.
This research has made use of NASA's Astrophysics Data System.
This research has made use of the NASA/IPAC Infrared Science Archive, which is operated by the Jet Propulsion Laboratory,
California Institute of Technology, under contract with the National Aeronautics and Space Administration.
This publication has made use of molecular line data from the Boston University-FCRAO Galactic Ring Survey (GRS).
The GRS is a joint project of Boston University and Five College Radio Astronomy Observatory,
funded by the National Science Foundation under grants AST-9800334, AST-0098562, \& AST-0100793.

\end{acknowledgements}


\bibliographystyle{aa}
\bibliography{sidorin_etal_2013_n107}

%
%

\Online

\onecolumn
\setlength\LTcapwidth{\hsize}

\begin{appendix}

\section{List of Molecular Clumps\label{sec_list_of_molecular_clumps}}

\begin{longtable}{lrrrrrrrrrrrrr}
  
  \caption{Molecular clumps associated with bubble N107.
    First column gives a number identifying a clump in our list.
    The meaning of other columns is explained in sec.~\ref{sec_analysis_of_molecular_clumps}.
    This is the full list expanding tab.~\ref{tab_clumps}.
    \label{tab_clumps_full}
    }\\
    \hline\hline
    $N$ &
    $\frac{l_\mathrm{peak}}{\mathrm{deg}}$ &
    $\frac{b_\mathrm{peak}}{\mathrm{deg}}$ &
    $\frac{v_\mathrm{peak}}{\mathrm{km\:s^{-1}}}$ &
    $\frac{T_{\mathrm{peak}}}{\mathrm{K}}$ &
    $\frac{R}{\mathrm{pc}}$ &
    $\frac{\mathrm{FWHM}_{l}}{\mathrm{pc}}$ &
    $\frac{\mathrm{FWHM}_{b}}{\mathrm{pc}}$ &
    $\frac{\mathrm{FWHM}_{v}}{\mathrm{km\:s^{-1}}}$ &
    $\frac{M}{M_\odot}$ &
    $\frac{\Delta M}{M_\odot}$ &
    $\frac{M_{\mathrm{vir}}}{M_\odot}$ &
    $\frac{\Delta M_{\mathrm{vir}}}{M_\odot}$ &
    $\frac{M}{M_{\mathrm{vir}}}$
    \\
    \hline
    \endfirsthead
  
  \caption{(continued)}\\
    \hline\hline
    $N$ &
    $\frac{l_\mathrm{peak}}{\mathrm{deg}}$ &
    $\frac{b_\mathrm{peak}}{\mathrm{deg}}$ &
    $\frac{v_\mathrm{peak}}{\mathrm{km\:s^{-1}}}$ &
    $\frac{T_{\mathrm{peak}}}{\mathrm{K}}$ &
    $\frac{R}{\mathrm{pc}}$ &
    $\frac{\mathrm{FWHM}_{l}}{\mathrm{pc}}$ &
    $\frac{\mathrm{FWHM}_{b}}{\mathrm{pc}}$ &
    $\frac{\mathrm{FWHM}_{v}}{\mathrm{km\:s^{-1}}}$ &
    $\frac{M}{M_\odot}$ &
    $\frac{\Delta M}{M_\odot}$ &
    $\frac{M_{\mathrm{vir}}}{M_\odot}$ &
    $\frac{\Delta M_{\mathrm{vir}}}{M_\odot}$ &
    $\frac{M}{M_{\mathrm{vir}}}$
    \\
    \hline
    \endhead
    
    \hline
    \multicolumn{14}{c}{(continuing on the next page)} \\
    \hline
    \endfoot
    
    \hline
    \endlastfoot
    
     1 & 50.919 &  0.241 &  41.89 &    7.0 &   2.2 &   2.45 &   1.62 &   1.84 & 1037.1 &    5.6 &  1409.9 &    324.8 &    0.736 \\
     2 & 50.840 &  0.266 &  44.01 &    5.8 &   4.0 &   3.48 &   3.40 &   2.83 & 4082.1 &   11.5 &  6050.2 &    909.0 &    0.675 \\
     3 & 50.753 & -0.029 &  45.08 &    4.7 &   3.2 &   3.38 &   2.52 &   4.63 & 3119.8 &   10.8 & 12847.9 &   1180.5 &    0.243 \\
     4 & 51.073 &  0.204 &  45.93 &    4.7 &   1.2 &   1.41 &   0.99 &   2.57 &  351.1 &    3.6 &  1494.4 &    247.5 &    0.235 \\
     5 & 50.766 &  0.081 &  45.50 &    4.6 &   3.3 &   3.66 &   2.21 &   3.92 & 3469.1 &   10.9 &  9622.4 &   1044.0 &    0.361 \\
     6 & 50.938 & -0.072 &  42.31 &    4.6 &   1.8 &   3.06 &   1.17 &   1.70 &  635.2 &    4.8 &  1010.1 &    252.4 &    0.629 \\
     7 & 50.778 &  0.192 &  42.74 &    4.5 &   5.0 &   5.15 &   5.07 &   2.76 & 7365.5 &   15.9 &  7259.4 &   1120.0 &    1.015 \\
     8 & 50.815 &  0.020 &  42.10 &    4.5 &   2.3 &   1.59 &   2.60 &   3.91 & 1206.7 &    6.8 &  6597.7 &    716.4 &    0.183 \\
     9 & 50.803 & -0.029 &  44.01 &    4.4 &   2.3 &   2.05 &   2.11 &   2.26 & 1059.1 &    6.1 &  2272.7 &    428.0 &    0.466 \\
    10 & 50.999 &  0.247 &  43.59 &    3.6 &   1.4 &   2.08 &   1.08 &   2.26 &  276.5 &    3.5 &  1343.3 &    252.2 &    0.206 \\
    11 & 50.858 & -0.048 &  42.95 &    3.3 &   1.2 &   1.72 &   0.86 &   1.55 &  169.3 &    2.8 &   560.8 &    154.1 &    0.302 \\
    12 & 50.796 &  0.124 &  44.01 &    2.5 &   1.4 &   1.27 &   1.41 &   1.44 &  149.6 &    2.7 &   554.4 &    164.0 &    0.270 \\
    13 & 50.784 &  0.131 &  46.56 &    2.3 &   0.9 &   0.87 &   1.09 &   0.96 &   45.3 &    1.5 &   152.0 &     67.4 &    0.298 \\
    14 & 50.870 & -0.152 &  40.83 &    2.1 &   1.0 &   1.22 &   1.06 &   1.42 &   60.9 &    1.7 &   366.1 &    109.3 &    0.166 \\
    15 & 50.969 & -0.128 &  39.34 &    2.0 &   0.8 &   1.01 &   0.87 &   0.58 &   26.3 &    1.2 &    53.4 &     39.3 &    0.493 \\
    16 & 50.883 &  0.297 &  43.80 &    1.9 &   0.5 &   0.88 &   0.46 &   0.47 &    8.4 &    0.7 &    20.0 &     18.3 &    0.417 \\
    17 & 50.913 &  0.180 &  46.35 &    1.9 &   0.9 &   1.61 &   0.93 &   1.09 &   48.5 &    1.6 &   197.8 &     77.0 &    0.245 \\
    18 & 50.913 &  0.155 &  41.68 &    1.8 &   0.6 &   0.84 &   0.67 &   0.46 &   11.7 &    0.8 &    25.2 &     23.1 &    0.465 \\
    19 & 50.864 &  0.210 &  45.29 &    1.8 &   0.5 &   0.68 &   0.52 &   1.11 &   15.6 &    0.9 &   114.0 &     43.7 &    0.137 \\
    20 & 50.833 &  0.210 &  40.19 &    1.7 &   1.0 &   1.36 &   1.06 &   0.81 &   26.3 &    1.2 &   120.1 &     63.4 &    0.219 \\
    21 & 50.833 &  0.057 &  42.95 &    1.7 &   0.5 &   0.73 &   0.76 &   0.21 &    5.0 &    0.5 &     4.6 &      9.2 &    1.088 \\
    22 & 50.864 &  0.260 &  41.04 &    1.7 &   0.4 &   0.73 &   0.50 &   0.41 &    5.8 &    0.5 &    14.2 &     14.6 &    0.411 \\
    23 & 50.987 & -0.171 &  41.46 &    1.6 &   0.5 &   0.96 &   0.58 &   0.21 &    4.0 &    0.5 &     4.2 &      8.4 &    0.950 \\
    24 & 50.956 & -0.122 &  40.40 &    1.6 &   0.8 &   1.27 &   1.06 &   0.79 &   22.7 &    1.1 &    99.3 &     53.6 &    0.228 \\
    25 & 50.790 &  0.014 &  38.91 &    1.6 &   0.6 &   0.85 &   0.79 &   0.30 &    6.2 &    0.6 &     9.7 &     13.8 &    0.639 \\
    26 & 50.870 & -0.042 &  39.55 &    1.6 &   0.8 &   1.02 &   1.00 &   0.59 &   17.8 &    1.0 &    54.8 &     39.2 &    0.326 \\
    27 & 50.735 &  0.051 &  39.76 &    1.6 &   0.6 &   1.02 &   0.59 &   0.41 &    8.7 &    0.7 &    19.3 &     20.2 &    0.449 \\
    28 & 50.846 &  0.192 &  39.13 &    1.5 &   0.4 &   0.79 &   0.57 &   0.31 &    4.4 &    0.5 &     8.0 &     11.0 &    0.548 \\
    29 & 50.833 & -0.066 &  41.68 &    1.5 &   0.5 &   0.95 &   0.57 &   0.27 &    4.6 &    0.5 &     7.0 &     10.8 &    0.661 \\
    30 & 50.889 & -0.036 &  39.76 &    1.5 &   0.4 &   0.59 &   0.55 &   0.32 &    3.8 &    0.5 &     7.4 &      9.8 &    0.519 \\
    31 & 51.061 & -0.042 &  45.71 &    1.5 &   0.5 &   0.89 &   0.49 &   0.48 &    5.5 &    0.5 &    21.5 &     19.0 &    0.255 \\
    32 & 50.981 &  0.026 &  43.80 &    1.5 &   0.5 &   0.98 &   0.59 &   0.21 &    4.0 &    0.5 &     4.2 &      8.4 &    0.955 \\
    33 & 50.778 &  0.032 &  46.78 &    1.5 &   0.6 &   0.96 &   0.56 &   0.48 &    9.6 &    0.7 &    24.9 &     22.2 &    0.383 \\
    34 & 50.864 &  0.087 &  42.31 &    1.4 &   0.5 &   0.77 &   0.76 &   0.21 &    4.2 &    0.5 &     4.2 &      8.4 &    0.992 \\
    35 & 50.981 & -0.152 &  46.14 &    1.4 &   0.5 &   0.80 &   0.59 &   0.21 &    3.8 &    0.5 &     4.2 &      8.4 &    0.899 \\
    36 & 50.987 &  0.038 &  46.99 &    1.4 &   0.4 &   0.53 &   0.81 &   0.33 &    3.8 &    0.5 &     8.8 &     11.5 &    0.426 \\
    37 & 50.864 &  0.192 &  42.10 &    1.4 &   0.5 &   0.70 &   0.71 &   0.61 &    8.0 &    0.7 &    38.3 &     26.5 &    0.210 \\
    38 & 51.006 &  0.247 &  45.50 &    1.4 &   0.3 &   0.59 &   0.39 &   0.42 &    3.8 &    0.5 &    10.3 &     10.5 &    0.370 \\
    39 & 50.969 &  0.327 &  46.78 &    1.3 &   0.6 &   1.02 &   0.59 &   0.30 &    5.6 &    0.6 &    10.2 &     14.2 &    0.554 \\
    40 & 50.956 & -0.109 &  46.35 &    1.3 &   0.5 &   0.52 &   1.10 &   0.21 &    3.5 &    0.5 &     4.2 &      8.4 &    0.844 \\
    41 & 50.956 & -0.091 &  46.99 &    1.3 &   0.5 &   1.06 &   0.57 &   0.21 &    4.4 &    0.5 &     4.6 &      9.2 &    0.964 \\
    42 & 50.969 &  0.327 &  41.89 &    1.3 &   0.4 &   0.73 &   0.50 &   0.32 &    4.6 &    0.5 &     8.5 &     11.3 &    0.538 \\
    43 & 50.852 & -0.017 &  45.29 &    1.3 &   0.4 &   0.83 &   0.53 &   0.29 &    3.6 &    0.5 &     6.9 &     10.2 &    0.521 \\
    44 & 50.741 &  0.051 &  42.10 &    1.3 &   0.5 &   0.69 &   0.78 &   0.21 &    3.7 &    0.5 &     4.2 &      8.4 &    0.892 \\
    45 & 50.821 &  0.241 &  39.55 &    1.3 &   0.6 &   0.59 &   1.11 &   0.33 &    7.8 &    0.7 &    11.8 &     15.3 &    0.665 \\
    46 & 50.815 & -0.066 &  42.95 &    1.2 &   0.5 &   1.15 &   0.56 &   0.40 &    5.7 &    0.6 &    16.0 &     17.1 &    0.356 \\
    47 & 51.221 &  0.081 &  46.78 &    1.2 &   0.5 &   0.71 &   0.71 &   0.32 &    5.6 &    0.6 &    10.6 &     13.9 &    0.526 \\
    48 & 51.067 &  0.149 &  45.93 &    1.2 &   0.4 &   0.53 &   0.78 &   0.29 &    3.5 &    0.5 &     6.9 &     10.1 &    0.513 \\
    49 & 50.796 &  0.223 &  41.04 &    1.2 &   0.4 &   1.01 &   0.39 &   0.29 &    3.4 &    0.5 &     7.0 &     10.2 &    0.485 \\
    
\end{longtable}

\end{appendix}

%
                            \end{document}